\documentclass[%
onecolumn,
superscriptaddress,
nofootinbib,
 amsmath,amssymb,
 aps]{revtex4-2}

\usepackage{graphicx}
\usepackage{tensor}
\usepackage{siunitx}
\usepackage{xcolor} 
\usepackage[normalem]{ulem} 
\usepackage{lipsum} 

\usepackage[pdfborder={0 0 0}, plainpages=false, hyperfigures=true]{hyperref}

\newcommand{\beq}{\begin{equation}}
\newcommand{\eeq}{\end{equation}}
\newcommand{\beqn}{\begin{eqnarray}}
\newcommand{\eeqn}{\end{eqnarray}}

\newcommand{\lo}{\mathrel{\raise.3ex\hbox{$<$}\mkern-14mu
    \lower0.6ex\hbox{$\sim$}}}
\newcommand{\go}{\mathrel{\raise.3ex\hbox{$>$}\mkern-14mu
    \lower0.6ex\hbox{$\sim$}}}

\newcommand{\UNH}{\affiliation{Department of Physics \& Astronomy, University of New Hampshire, 9 Library Way, Durham NH 03824, USA}}

\begin{document}
\title{Neutrinos in colliding neutron stars and black holes}
\author{Francois Foucart}\UNH

\begin{abstract}
	In this chapter, we provide an overview of the physics of colliding black holes and neutron stars and of the impact of neutrinos on these systems. Observations of colliding neutron stars play an important role in nuclear astrophysics today. They allow us to study the properties of cold nuclear matter and the origin of many heavy elements (gold, platinum, uranium). We show that neutrinos significantly impact the observable signals powered by these events as well as the outcome of nucleosynthesis in the matter that they eject into the surrounding intergalactic medium.
\end{abstract}

\maketitle

\section{Introduction}\label{intro}

Compact objects such as black holes and neutron stars provide us with a remarkable laboratory to test the laws of physics in extreme environments that cannot be reproduced by scientific experiments on Earth -- or, really, anywhere else in the known Universe. 
Black holes and neutron stars are the end point of the evolution of massive stars ($M\gtrsim 8M_\odot$, with $M_\odot$ the mass of the Sun), after these stars run out of nuclear fuel to burn and collapse under their own gravitational fields. Neutron stars have expected radii of only $\sim (10-14)\,{\rm km}$, roughly the size of a small city, but masses of $\sim (1-2)M_\odot$. Their cores reach densities $\gtrsim 10^{15}\,{\rm g/cm^3}$, higher than that of atomic nuclei. We know very little about the behavior of matter at such densities, and that behavior remains an important open question in nuclear physics today (see e.g.~\cite{Lattimer:2021emm} for more information on neutron stars and the properties of high-density matter). Black holes, of course, are even denser than neutron stars. Mathematically, they are point masses surrounded by a region from which nothing can escape their gravitational pull\footnote{Or a little more accurately, in the language of general relativity, the strong spacetime curvature of the black hole creates a region from which no information can escape}. The boundary of that region is what we call the {\it event horizon} of the black hole. Black holes of any mass can, in theory, exist. Those formed in the collapse of massive stars range from a few times the mass of the sun to tens of solar masses. For a black hole of mass $M_{\rm BH}$, the event horizon has a size of $\sim (1.5-3)\left(\frac{M_{\rm BH}}{M_\odot}\right)\,{\rm km}$ (the exact size depends on how fast the black hole is rotating)~\cite{Bardeen:1972fi}. A typical black hole of mass $M_{\rm BH}\sim (5-10)M_\odot$~\cite{2010ApJ...725.1918O,Farr:2010tu} will thus have a size comparable to that of a neutron star. With so much mass within a very small volume, black holes and neutron stars offer us unparalled opportunities to study the laws of gravity as well as, in the case of neutron stars, the properties of extremely dense matter.
 
A relatively rare but particularly interesting astrophysical event is the collision of two compact objects, and their eventual merger into a more massive black hole or neutron star. The existence of these merger events is a consequence of Einstein's theory of general relativity, which predicts that orbiting objects slowly loose energy through the emission of gravitational waves. This energy loss increases with the mass and acceleration of the orbiting bodies, and causes their separation to slowly decrease over time. Energy losses from gravitational wave emission are negligible for most astrophysical system (e.g. the Earth-Moon system is practically unaffected by gravitational wave emission), but can be quite significant for orbiting compact objects. Consider that two compact objects can orbit each other at a distance of $\sim 30\,{\rm km}$ without direct contact, at which point their orbital velocity is about $30\%$ of the speed of light! A lot of things have to go right for them to reach that state, either through the slow gravitational-wave driven decay of their orbits for systems formed from high-mass stellar binaries, or from gravitational capture for systems formed in a dense stellar environment. Current estimates indicate that in a Milky-Way-like galaxy, compact object mergers occur only once every $10^{4-5}$ years~\cite{KAGRA:2021vkt}. Yet the collision at a significant fraction of the speed of light of solar-mass objects releases so much energy that we can observe these mergers billions of light-year away. The first black hole binary merger was observers in 2015~\cite{LIGOScientific:2016aoc}, while the first binary neutron star was observed in 2017~\cite{LIGOScientific:2017vwq} and two mixed black hole-neutron star system in 2020~\cite{LIGOScientific:2021qlt}. Today, more than a hundred binary black hole systems have been detected, as well as two confirmed neutron star binaries and $\sim 5$ mixed binaries (a number of events of marginal significance make it hard to agree on an exact number of detections)~\cite{KAGRA:2021vkt}.

\begin{figure}[h]
	\centering
	\includegraphics[width=0.3\linewidth]{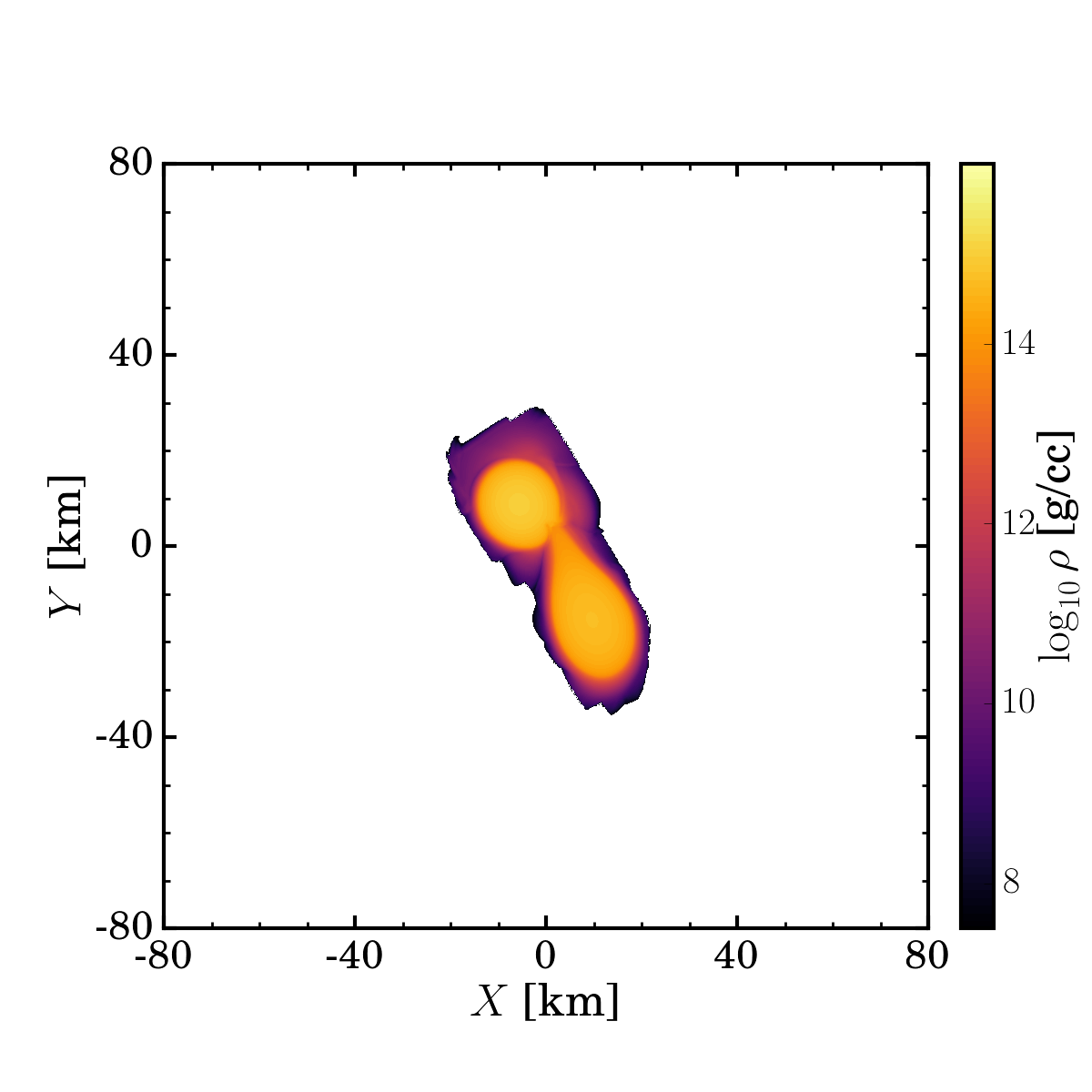}
	\includegraphics[width=0.3\linewidth]{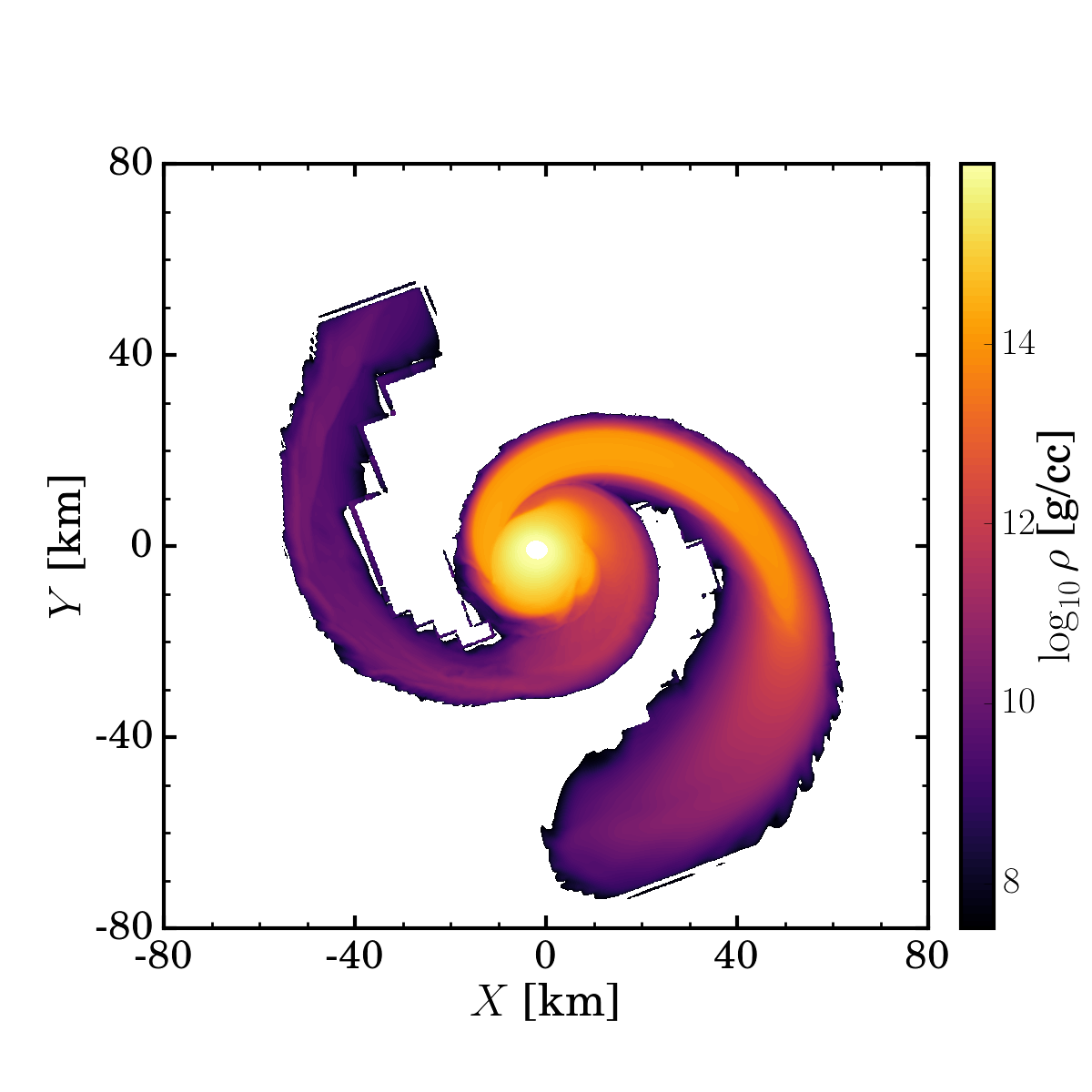}
	\includegraphics[width=0.3\linewidth]{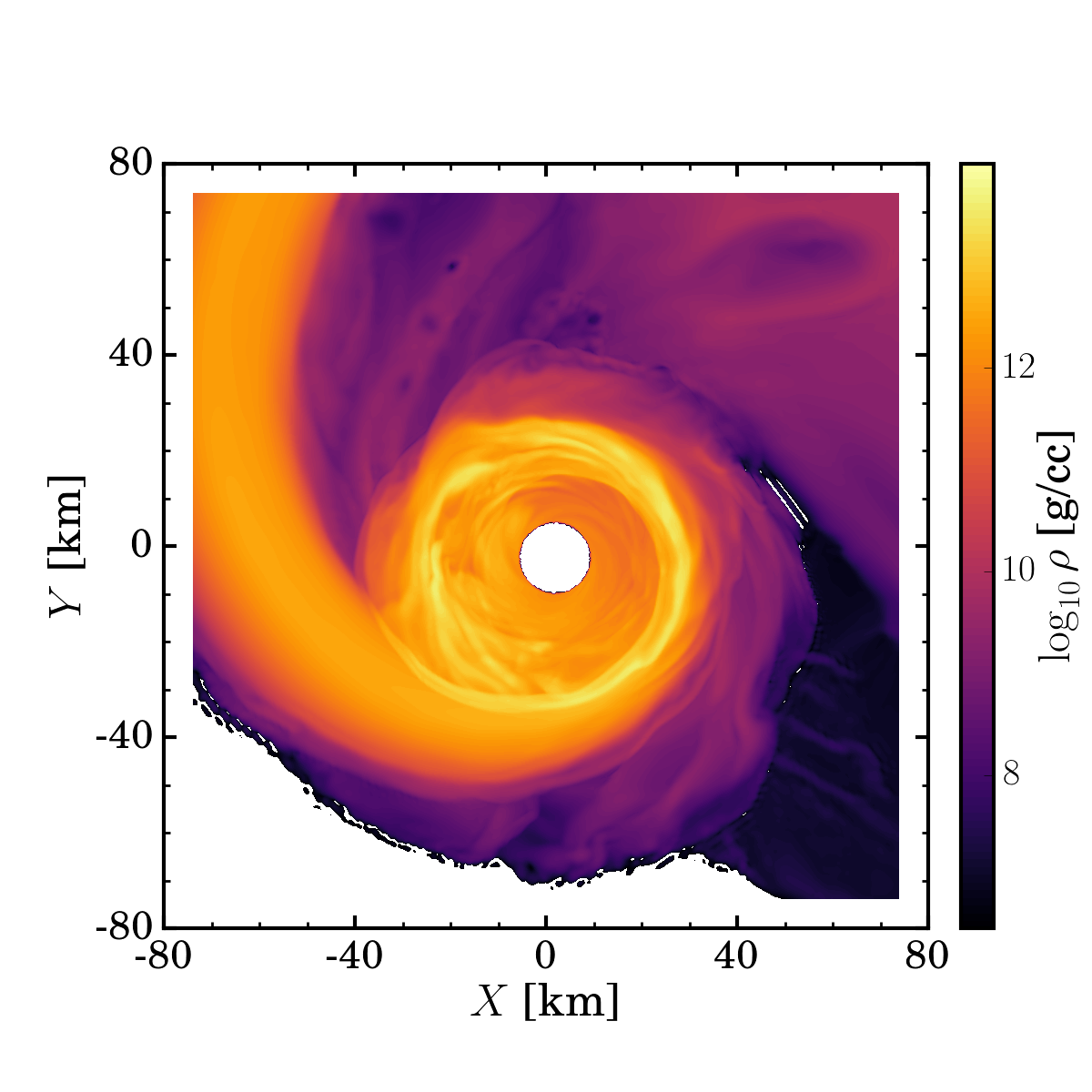}
	\caption{Collision of two neutron stars as simulated with the SpEC code. We show the matter density in the equatorial plane of the binary. {\it Left:} The two neutron star get into contact, with the lower mass neutron star (bottom) tidally distorted by the higher mass neutron star. {\it Middle:} The two cores form a single central object. Matter from the neutron star is ejected in cold tidal tails, mainly from the lower mass neutron star. {\it Right:} The central object has collapsed to a black hole (white circle), while bound matter from the tidal tails is now forming an accretion disk around the black hole. The entire merger process shown here lasts less than $4\,{\rm ms}$. Images adapted from~\cite{Foucart:2022kon}.}
	\label{fig:bns}
\end{figure}

An overview of the collision process for neutron star binaries is shown on Fig.~\ref{fig:bns}. As the cores of the neutron stars merge, the lower mass neutron star is strongly distorted by the gravitational field of its companion. A small part of the neutron star is ejected into the surrounding interstellar medium, while more matter remains bound in an accretion disk around the remnant. The collision itself results in additional ejection of hot, shocked material in the interstellar medium. As for the merging cores, they either collapse to a black hole (for high-mass systems), or form a more massive neutron star (for low-mass systems).This is an extremely fast process: the merger itself lasts only a few milliseconds! The collapse of the neutron star to a black hole can immediately follow the merger (i.e. on milliseconds timescales), or for intermediate mass system be delayed (on basically any possible time scale). In the few seconds following the merger, most of the matter in the disk is more slowly accreted by the black hole or neutron star remnant, while the rest is ejected in the form of disk winds. These various channels for matter ejection might play a surprisingly important role in the formation of many heavy elements observed today. About half of the elements heavier than iron, including most of the gold, platinum, and uranium in the Universe, are expected to be formed through nuclear reactions in neutron-rich environments. The exact production site of these elements remains unknown, but the matter ejected by neutron star mergers is a likely candidate~\cite{Kasen:2017sxr,2017Natur.551...67P}. The radioactive decay of the heavy nuclei formed in the ejected matter also results in the emission of an observable optical/infrared signal days to weeks after the merger, called a {\it kilonova}~\cite{1976ApJ...210..549L,2011ApJ...736L..21R}. We will discuss these processes in more detail later in this chapter. Black hole-neutron star mergers can similarly eject material and form black hole-disk systems through the disruption of the neutron star by its black hole companion. This is however only possible for relatively low mass or rapidly spinning black holes; in other systems, the neutron star plunges into the black hole before being disrupted. More detailed discussions of our current understanding of neutron star mergers can be found e.g. in~\cite{Baiotti:2016qnr,Burns:2019byj,Foucart:2020ats,Radice:2020ddv,Kyutoku:2021icp}.

What about neutrinos? Immediately after merger, the remnant and surrounding disk reach temperatures of $10^{11-12}\,{\rm K}$. The accretion disk has a typical density of $10^{10-12}\,{\rm g/cm^3}$. Under these conditions, photons are trapped in the dense matter, unable to escape. In fact, inside a neutron star remnant, even neutrinos are trapped! Neutrinos may however escape from the surface of the neutron star remnant and/or from the accretion disk, and are produced in copious quantities in those regions. Neutrino emission thus becomes the dominant cooling mechanism for the remnant. Weak interations $n+e^+\leftrightarrow p + \bar\nu_e$ and $p+e^- \leftrightarrow n+ \nu_e$ additionally allow for the conversion of neutrons into protons, and vice-versa. As a result, neutrino-matter interactions impact the neutron-richness of the post-merger remnant and of the matter ejected during and after the collision. This change in neutron-richness has significant effects on the outcome of element formation in the ejected matter, and on the observable properties of kilonovae~\cite{Barnes:2013wka}. Overall, it is impossible to truly understand the role that neutron star mergers play in the production of heavy elements or to analyze the electromagnetic signals powered by these mergers without a careful study of the role of neutrinos in the evolution of their post-merger remnants. More advanced neutrino physics may also impact observable properties of neutron star mergers. Collective neutrino oscillations could for example play a role in setting the properties of the matter ejected by compact object mergers.

In this chapter, we will begin with a brief overview of nuclear physics questions that are of particular importance for the study of merging neutron stars (Sec.~\ref{nucastro}). We will then discuss in more detail the role that neutrinos play in neutron star mergers (Sec.~\ref{nu}), before briefly reviewing the impact of neutrinos on the observable propeties of mergers (Sec.~\ref{obs}).

\section{Nuclear Astrophysics with neutron star mergers}\label{nucastro}

\subsection{Properties of dense matter}\label{nucastro:eos}

The properties of the cold, dense matter at the core of neutron stars is an important open question in nuclear physics today, as it is tightly tied to the unknown strength of the strong nuclear forces applied on nucleons (neutrons and protons) at close separation. Most of the matter inside of a neutron star is, as the name implies, very neutron rich ($>90\%$ neutrons, with some protons and electrons). Repulsive forces between nucleons are what allows the neutron star to support itself against gravitational collapse in that regime. At the highest densities that can be encountered at the center of the most massive neutron stars, particles including strange quarks can additionally be created, and we may even observe a  transition to an exotic phase\footnote{For example a transition from standard nuclear matter with quarks bound in hadrons to deconfined quarks, either in a homogeneous superconducting phase~\cite{RevModPhys.80.1455} or in inhomegeneous phases~\cite{Buballa:2020xaa}}. When or even whether this happens remains unclear. Interestingly, there is a direct connection between the size of a neutron star of a given mass and the pressure within its core. That pressure is itself directly related to the strength of nuclear interactions between nucleons, and to the composition of the core~\cite{Lattimer:2021emm}. The maximum mass that a neutron star can reach without collapsing to a black hole is impacted by the same nuclear physics unknowns. As a result, measurements of the sizes and masses of neutron stars have become an important tool in observational investigations of nuclear physics at high densities~\cite{GW170817-NSRadius,2017ApJL2041,Raaijmakers:2021uju,Miller:2021qha}. Observations of neutron stars are in that respect complemented by measurements of the internal structure of heavy nuclei such as lead in laboratories on Earth~\cite{Horowitz:2013wha,PhysRevLett.126.172503}, which is similarly impacted by the forces between multiple nucleons at close separation. 

As this chapter focuses mostly on neutrino physics, we will not cover the properties of dense matter in more detail but refer the interested reader to e.g.~\cite{Lattimer:2021emm} for more details. For our purpose, it will be enough to know that measuring neutron star masses and radii can be extremely valuable to improve our understanding of nuclear physics. In the context of neutron star mergers, the size of a neutron star determines how easy it is to disrupt, and thus how much matter is ejected in that disruption and how massive of an accretion disk will be formed around the post-merger remnant. The size of neutron stars also impacts their separation and, consequently, velocity when they collide. The mass of merging neutron stars determines whether they quickly collapse into a black hole, or can survive as a neutron star. Finally, both mass and radius impact the gravitational wave emission of neutron star binaries. While pre-merger gravitational wave emission is independent of any neutrino physics, the properties of the ejected matter and the resulting observable electromagnetic signals are definitely impacted by neutrinos. As a result, our ability to recover information about the properties of dense matter from electromagnetic emission powered by neutron star mergers depends in part on a sufficient understanding of neutrino physics in these systems.

\subsection{Formation of heavy elements: r-process nucleosynthesis}\label{nucastro:rprocess}

Understanding where, when, and how the various atomic elements observed in the Universe today were synthesized has long been an objective of nuclear astrophysicists. The production site of about half of the elements heavier than iron, in particular, remains an open question. Heavy elements are mainly produced through two distinct processes: the slow neutron capture process (s-process) and the rapid neutron capture process (r-process). In both cases, heavier nuclei are created by neutron capture onto lower mass nuclei, through e.g. 
\begin{equation}
X^A_Z + n \rightarrow X^{A+1}_Z + \gamma
\end{equation}
where $X^A_Z$ denotes a nucleus with $Z$ protons and $(A-Z)$ neutrons.
Unstable neutron-rich nuclei formed through these neutron captures then undergo beta decay 
\begin{equation}
X^A_Z \rightarrow X^A_{Z+1} + e^- + \bar\nu_e,
\end{equation}
creating more stable, less neutron-rich nuclei. The main difference between the r-process and s-process is the relative timescales between neutron captures and beta decays: in the s-process, neutron captures are slower than the beta decay timescale of nuclei close to the valley of stability. Nuclei thus remain close to that valley of stability (in terms of their number of protons and neutrons). This is known to happen in low mass stars towards the end of their life. In the r-process, on the other hand, neutron capture is much faster than beta decay, leading to the production of very neutron rich, highly unstable nuclei. If enough neutrons are present, neutron capture can eventually create heavy elements that are unstable to fission. The fission fragments can then capture more neutrons, continuing the r-process. The r-process is the only known pathway to naturally synthesize the heaviest elements found in nature, the actinides (e.g. U, Pu). The r-process naturally requires very neutron rich material like the matter ejected during and after neutron star mergers. While signs of r-process nucleosynthesis have been observed in the aftermath of at least one neutron star merger~\cite{Kasen:2017sxr,2017Natur.551...67P}, we do not at this point know whether neutron star mergers are sufficiently frequent and eject enough mass to explain the abundances of r-process elements observed in the Universe, or if another production site for the r-process is necessary. For recent reviews of the r-process and potential astrophysical production sites, see e.g.~\cite{Kajino:2019abv,Siegel:2022upa}

\begin{figure}[h]
	\centering
	\includegraphics[width=0.7\linewidth]{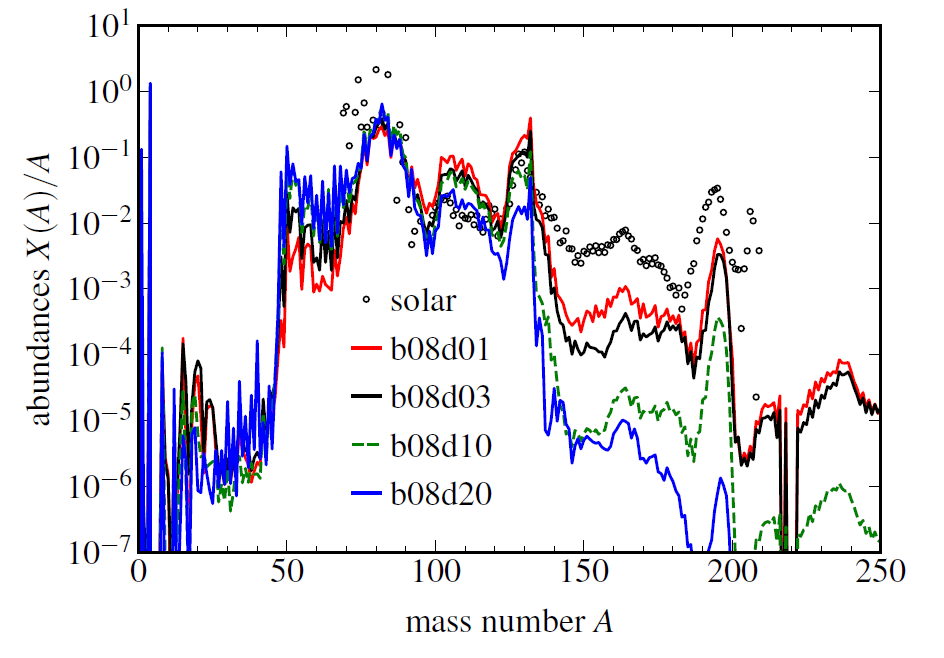}
	\caption{Relative abundances of elements synthesized in outflows from accretion disks produced in black hole-neutron star mergers. The dots show the inferred abundances of r-process elements in the solar system. The lines correspond to systems with different disk masses, which lead to different outflow compositions (parametrized by the electron fraction $Y_e$, as defined in Eq.~\ref{eq:ye}): more neutron-rich outflows produce more heavy elements (red curve, average $Y_e=0.28$); less neutron rich outflows produce only lower mass elements (blue curve, average $Y_e=0.33$). Figure reproduced from~\cite{Fernandez:2020oow}}
	\label{fig:rprocess}
\end{figure}

Neutrino matter interactions are crucial to determine the outcome of r-process nucleosynthesis. Which elements are produced at the end of the r-process and in what quantity is sensitive to the initial neutron-richness of the matter~\cite{Surman:2005kf,Lippuner2015} (see also Fig.~\ref{fig:rprocess}), and we will see that one of the main effect of neutrino-matter interactions in mergers is to change the neutron-richness of the ejected matter and post-merger remnant. In neutron star mergers, the composition of the matter is usually described through the net electron fraction $Y_e$, defined as
\begin{equation}
Y_e = \frac{n_{e^-}-n_{e^+}}{n_p + n_n}.
\label{eq:ye}
\end{equation}
As the fluid has no net charge, we have $n_{e^-}-n_{e^+}=n_p$ at least in regions of the remnant where most particles are $(p,n,e^+,e^-)$. In those regions, $Y_e=Y_p=(1-Y_n)$, with $Y_{p,n}$ the proton/neutron fraction, defined in the same was as $Y_e$ but replacing the numerator $(n_{e^-}-n_{e^+})$ with $n_{p,n}$. A low electron fraction thus corresponds to neutron-rich matter, and a high electron fraction to proton-rich matter.
For the typical properties of merger outflows, the formation of the heavier r-process elements ($A\gtrsim 140$) requires $Y_e\lesssim 0.25$\cite{Lippuner2015}. However, in such neutron-rich ejecta, the lower mass r-process elements are underproduced compared with astrophysical and solar system observations. Moderately neutron rich matter ($Y_e\sim 0.25-0.5$) on the other hand does not produce the higher mass elements that we observe in nature, but does produce the lower-mass r-process elements. If neutron star mergers are the main source of r-process elements in the Universe, we thus need both very neutron rich outflows and moderately neutron rich outflows. As the matter from the neutron star originally is $>90\%$ neutrons ($Y_e<0.1$), this is only possible if the outflows have significant interaction with neutrinos, making them less neutron rich. We should also note that the exact boundary between those two regimes vary with the properties of the outflows (temperature, velocity) and that smaller variations in the electron fraction of the outflows have a non-negligible impact on the exact abundance pattern of the elements created through r-process nucleosynthesis. This is particularly true for the abundance of actinides, which can vary significantly with the neutron-richness of the outflows, and is also observed to be quite variable in observations of low-metallicity stars~\cite{Holmbeck:2020qlp} -- i.e. stars formed so early in the evolution of their galaxies that their r-process elements are likely produced in a single event.

Overall, we thus note that the fraction of neutrons present in the matter ejected by neutron star mergers will be a crucial determinant of the outcome of r-process nucleosynthesis in these events, and that this fraction is largely set by neutrino-matter interactions in the ejecta.

\subsection{Kilonovae}\label{nucastro:kn}

An important observable whose properties will be in large part determined by neutrino-matter interactions is the kilonova signal powered by the matter ejected in compact object mergers. Within the first hours following the merger, that ejecta is dense and hot enough that optical/infrared photons are trapped within the ejected matter. The energy released by radioactive decays in the outflows is either deposited as heat in the ejecta, or lost to escaping neutrinos. As the ejected matter expands and cools, however, its opacity to photons eventually becomes small enough for photons to escape. The energy released by radioactive decays is then partially converted into a detectable optical/infrared signal, the kilonova~\cite{1976ApJ...210..549L,2011ApJ...736L..21R}.

Observations of kilonovae could play an important role in our understanding of the properties of dense matter, and of the outcome of r-process nucleosynthesis in the ejected matter. The kilonova signal depends on the mass, temperature, composition, and geometry of the ejected matter~\cite{Barnes:2013wka}. Its composition dependence is particularly interesting, due to its connection to r-process nucleosynthesis. We have already mentioned that more neutron rich outflows produce heavier r-process elements. An important consequence of this result is that only outflows with $Y_e\lesssim 0.25$ produce lanthanides ($57\leq Z \leq 71$) and actinides ($89\leq Z \leq 102$). These elements are of particular relevance because of their many bound-bound transitions at energies within the infrared/optical bands. Their presence in the outlfows thus delays the transition of the matter from optically thick to photons to optically thin. This leads to the kilonova signal becoming visible at a later time -- and typically at lower luminosity and with photons of higher wavelength. In the presence of lanthanides/actinides, kilonovae peak in the infrared on a $\sim$ week time scale instead of the optical on a $\sim$ day time scale. Pratically, this means that the observation of the color and duration of a kilonova provides us with very useful information about the outcome of r-process nucleosynthesis in the ejected matter. Because the mass and size of the original neutron stars impact the properties of the ejected matter, kilonovae can also in theory complement gravitational wave observations in our quest to constrain the properties of neutron stars. This is a difficult process, however. To recover that information, one needs a reliable map between the properties of the initial neutron stars and the eventual kilonova signal. This requires detailed predictions of the outflow properties, an understanding of the outcome of r-process nucleosynthesis, and accurate predictions for the opacity of the outflows to photons.  Such a map cannot be reliably produced at this time due in part to the difficulty of properly accounting for neutrino-matter interactions in merger simulations. Other important sources of uncertainty include our poor understanding of the properties of the very neutron-rich nuclei involved in the r-process (and of the reaction rates involved in the r-process), the limited accuracy of our predictions for the opacity of lanthanides and actinides to optical/infrared photons, as well as a limited ability to capture important small scale effects in the evolution of highly magnetized fluids in merger simulations. A significant amount of work is thus still required in order to be able to reliably use kilonovae to estimate the parameters of merging neutron stars, with neutrino-matter interactions playing a large role in this process.

\section{Neutrinos in neutron star mergers}\label{nu}

\subsection{Important reactions}\label{nu:reac}

At the temperatures and densities found in neutron star mergers, neutrinos mostly come into play through four broad classes of processes: charged-current reactions, pair production/annihilation, scatterings, and collective flavor oscillations. In the following sections, we provide a brief overview of how each reaction matters in the merger context. We note that in order of magnitude estimates provided for the neutrino mean-free paths, we completely ignore blocking factors (i.e. corrections to the interaction rate due to the Pauli exclusion principle, which prevents two fermions from being in the exact same quantum state) and the finite mass of the electrons. These estimates are thus very inaccurate in the highest density regions and in regions with temperature of $\sim 1\,{\rm MeV}$ or lower. They are however reasonable in the regions most important for the qualitative description of neutrino physics in mergers presented below. We refer the reader to~\cite{PhysRevD.29.1918,1985ApJS...58..771B,1999ApJ...517..859S} for more accurate calculations of reaction rates and to~\cite{Burrows:2004vq,Foucart:2022bth} for more detailed reviews of these processes.

\subsubsection{Charged current reactions}

Charged current reactions are by far the most important in the merger context, as they allow for the conversion of neutrons into protons, and vice-versa. We will mainly encounter charged current reactions involving electron-type (anti)neutrinos:
\begin{equation}
n + e^+ \leftrightarrow p+ \bar \nu_e;\,\, p + e^- \leftrightarrow n + \nu_e
\end{equation}
The standard neutron decay reaction $n\rightarrow p + e^- + \bar\nu_e$ is much slower than the above reactions in most regions of the merger. Charged current reactions involving muons and taus, e.g. $n + \mu^+ \leftrightarrow p+ \bar \nu_\mu$, are not as common in most of the post-merger remnant. This is due to the fact that the muon has a mass of $\sim 106\,{\rm MeV/c^2}$, while most of the remnant is at temperature $T\sim (1-50)\,{\rm MeV}$. Muon production is thus suppressed, except for a few regions at the interface between the colliding neutron stars where temperatures $\sim 100\,{\rm MeV}$ can be found. Recent results  indicate that muon charged-current reactions may not always be negligible~\cite{Loffredo:2022prq}, though they are certainly less important than electron charged-current reactions. Tau lepton production is completely negligible in neutron star mergers.

\begin{figure}[h]
	\centering
	\includegraphics[width=0.5\linewidth]{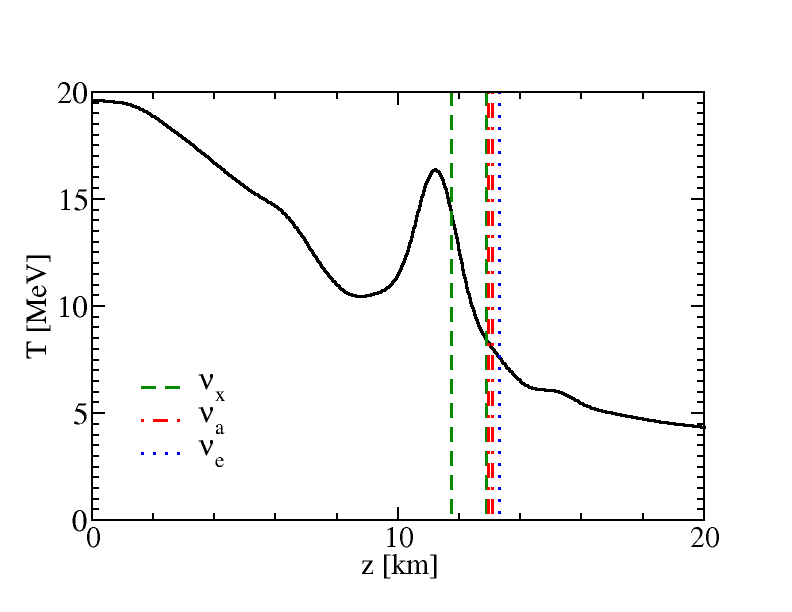}
	\caption{Temperature of the fluid and location of the surfaces where neutrino decouple from the fluid on a vertical line along the axis of rotation of a post-merger neutron star remnant, $10\,{\rm ms}$ after merger. For each species of neutrinos, two surfaces are shown: farther left is the surface where the optical depth to absorption is $\sim 2/3$, and farther right the surface where the optical depth to scattering is $\sim 2/3$. The two surfaces are nearly the same for $\nu_e$ neutrinos, very close for $\bar\nu_e$ (labeled as $\nu_a$ on the figure), and widely separated for $\nu_x$. Left of the first surface, neutrinos are in equilibrium with the fluid. In between the two surfaces, neutrinos random-walk through the fluid, on average slowly diffusing outward, without being absorbed. Right of the second surface, neutrinos are largely free-streaming. Due to the different temperatures on the decoupling surfaces, the energy of $\nu_x$ outside the neutron star is larger than that of $\bar\nu_e$, which is itself higher than that of $\nu_e$. Figure reproduced from~\cite{Foucart:2016rxm}.}
	\label{fig:neutrinosphere}
\end{figure}

To understand the role of charged current reactions in mergers, it is useful to remember how the reaction rates approximately scales. For example, the absorption opacity (i.e. the inverse mean-free path) due to the reaction $n + \nu_e \rightarrow p + e^-$ is, at the order of magnitude level~\cite{1985ApJS...58..771B,Burrows:2004vq,Foucart:2022bth},
\begin{equation}
\kappa_{n\nu_e} \approx 1.38 \sigma_0 n_n \left(\frac{\epsilon}{m_e c^2}\right)^2
\end{equation}
with $n_n$ the number density of neutrons, $\epsilon$ the energy of the incoming neutrino, $m_e$ the mass of an electron, and $\sigma_0 = 1.705\times 10^{-44}\,{\rm cm^{2}}$. We can rewrite this as
\begin{equation}
\kappa_{n\nu_e} \approx \left(140\,{\rm km^{-1}}\right) \left(\frac{\epsilon}{10\,{\rm MeV}}\right)^2 \frac{n_n}{0.15\,{\rm fm^{-3}}}
\end{equation}
if we scale the neutron number density by the nuclear saturation density $0.15\,{\rm fm^{-3}}$ ($\sim 2.5\times 10^{14}\,{\rm g/cm^3}$ in rest mass density) and the neutrino energy by a typical value of $10\,{\rm MeV}$ for neutrinos emitted in neutron star mergers. We see that $\kappa_{n\nu_e}\approx 1\,{\rm km^{-1}}$ at about $1\%$ of the nuclear saturation density, i.e. at that density neutrinos will have a mean-free path of about $1\,{\rm km}$. At larger densities, neutrinos will be effectively trapped in the fluid (constantly reabsorbed and reemitted). At lower densities, they will largely decouple from the fluid. A similar relationship can be derived for the absorption of $\bar\nu_e$ through $\bar \nu_e + p \rightarrow n + e^+$,
\begin{equation}
\kappa_{p\bar\nu_e} \approx \left(140\,{\rm km^{-1}}\right) \left(\frac{\epsilon}{10\,{\rm MeV}}\right)^2 \frac{n_p}{0.15\,{\rm fm^{-3}}}
\end{equation}
The main difference will be that inside the neutron star, $n_p \lesssim 0.1 n_n$, and thus $\bar\nu_e$ will decouple from the fluid at a total baryon density higher than for $\nu_e$ neutrinos. This is slightly compensated by the fact that temperatures are higher at higher densities in post-merger remnant, and thus the energy of $\bar\nu_e$ as they decouple from the fluid is higher than that of $\nu_e$. We will see below that scattering onto neutrons also reduces the distance between the decoupling region of $\nu_e$ and $\bar\nu_e$, but without completely erasing that difference. The location of the decoupling surfaces is shown on Fig.~\ref{fig:neutrinosphere} along the rotation axis of a post-merger neutron star remnant.

Within a neutron star remnant, we thus transition from a ``trapped'' region close to the core to a ``free-streaming'' region on the surface. Within a post-merger disk, typical initial densities are $10^{-4}-10^{-2}$ of the nuclear saturation density and temperature $\sim (1-10)\,{\rm MeV}$. At their highest density, post-merger accretion disks have scale heights of $O(10\,{\rm km})$ and their optical depth to electron-type neutrinos is thus initially up to $O(1)$. Electron neutrinos are not trapped in the disk, but they have a significant probability of being absorbed by the fluid before escaping the disk. Muon and tau neutrinos are more rarely absorbed at those lower densities and temperature, as the neutrinos propagating through the disk do not have enough energy to produce muon and tau leptons.

\subsubsection{Pair processes}\label{sec:pairs}

The second broad class of reactions important in the merger context are pair processes such as electron-positron annihilation (and the inverse process of neutrino-antineutrino annihilation)
\begin{equation}
e^-+ e^+ \leftrightarrow \nu + \bar \nu
\end{equation}
and nucleon-nucleon Bremsstrahlung. Here, the neutrinos can be of any flavor, though the exact reaction rate varies with neutrino flavor. The forward reaction depends mostly of the availability of electron-positron pairs, and the reaction rate will thus increase steeply with temperature (the energy emission rate scales as $T^9$). For the typical conditions encountered in mergers, pair production of $\nu_e\bar\nu_e$ is typically slower than emission through charged current reactions. For production of muon and tau neutrinos, given the low rate of charged current reactions, this will often be the dominant process. As a result, heavy-lepton neutrinos decouple from the fluid at higher density and temperature than electron type neutrinos within post-merger neutron stars (see Fig.~\ref{fig:neutrinosphere}), and are rarely absorbed within post-merger accretion disks.

The inverse reaction ($\nu\bar\nu$ annihilation) depends on the availability of neutrino-antineutrino pairs. We additionally note that the cross section of the inverse reaction is proportional to $(1-\cos\theta)^2$, with $\theta$ the angle between the direction of propagation of the neutrino and antineutrino~\cite{1999ApJ...517..859S}. Significant annihilation rates in free-streaming regions thus require neutrinos and antineutrinos coming from different regions of the remnant. We will see later that this may happen in the polar regions along the rotation axis of a post-merger remnant.

\subsubsection{Scattering}

All species of neutrinos additionally interact with the matter through scattering onto neutrons, protons, electrons and nuclei. The scattering opacity (inverse mean-free path) for collisions on neutrons and protons can be approximated as~\cite{1985ApJS...58..771B,Burrows:2004vq,Foucart:2022bth}
\begin{equation}
\kappa_{s,n} \approx \left(33\,{\rm km^{-1}}\right)\left(\frac{\epsilon}{10\,{\rm MeV}}\right)^2\frac{n_n}{0.15\,{\rm cm^{-3}}};\,\,
\kappa_{s,p} \approx \left(28\,{\rm km^{-1}}\right)\left(\frac{\epsilon}{10\,{\rm MeV}}\right)^2\frac{n_p}{0.15\,{\rm cm^{-3}}}.
\end{equation}
These scattering events are nearly elastic (the energy of the neutrino does not vary much in the center-of-mass frame for $\sim 10\,{\rm MeV}$ neutrinos scattering on baryons with rest mass of $\sim 1\,{\rm GeV}$), and nearly isotropic ($\sim (10-20)\%$ changes in the differential cross-section as a function of the direction of propagation of the outgoing neutrino). In post-merger remnants, scattering onto neutrons is clearly the dominant effect, as $n_n \gg n_p$. We also see that for $\nu_e$, the scattering cross-section is smaller than the cross section of charged-current reaction; scattering events will not modify much where neutrinos are trapped or free-streaming. For $\bar\nu_e$, the scattering opacity can be slightly larger than the charged-current opacity, on the other hand. In a neutron star, this will contribute to bringing the decoupling surface between neutrinos and the fluid to lower densities: neutrinos undergoing scattering events effectively random-walk through the remnant, giving them more time to potentially be absorbed by charged-current reactions. For heavy-lepton neutrinos, there will be a relatively large region close to the surface of neutron stars where the mean-free path of neutrinos to scattering events is short, but the mean free path to absorption is large. In those regions, neutrinos slowly diffuse through the remnant, with limited changes to their spectrum. These effects are all visible on Fig.~\ref{fig:neutrinosphere}. In accretion disks, all types of neutrinos may see scattering optical depths of $O(1)$. At the order-of-magnitude level used in this section, this does not significantly impact our predictions -- but this shows that scattering cannot be neglected when simulating the effect of neutrinos in post-merger remnants. Scattering onto nuclei is generally subdominant, due to the relatively low number of nuclei in the post-merger remnant (compared to free protons and neutrons) -- at least at densities at which neutrino-matter interactions are most relevant. 

Scattering on electrons is not as frequent as scattering on neutrons in the regions of parameter space encountered in post-merger remnant, but it may have a mean-free-path similar to that of pair annihilations~\cite{1985ApJS...58..771B,Burrows:2004vq,Foucart:2022bth}. As electrons are relativistic in merger remnants scattering on electrons is very inelastic: the energy of both the electrons and the neutrinos can change significantly after a scattering event. As a result, scattering events could impact the spectrum of heavy-lepton neutrinos escaping post-merger neutron stars. Due to the high cost of including inelastic scattering in simulations of neutron star mergers, the exact effect of inelastric scattering on electrons remains unknown today.

\subsection{Neutrino Flavor Conversion}

In addition to their interactions with matter, the distribution of neutrinos in neutron star mergers may be impacted by flavor conversion. The fact that the mass eigenstates of neutrinos are a superposition of all three neutrino flavors is well known, and results in vacuum oscillations as well as flavor oscillations within slowly varying matter density profiles (MSW mechanism) that were, for example, crucial to understand the neutrino signal coming from the Sun~\cite{Bahcall:1976zz,Wolfenstein:1977ue,Bilenky:1978nj,Mikheyev:1985zog}. Flavor oscillations over length scales much larger than the size of the remnant neutron stars can modify the luminosity of neutrinos of different flavor as observed from Earth. Unfortunately, for the neutrinos emitted from the neutron star remnant and disk at least, such direct observations are unlikely to be possible lacking an extremely lucky close-by event (i.e. within our galaxy or local group). For example, water detectors such as Hyper-Kamiokande may need to monitor thousands of mergers to detect a single neutrino~\cite{Kyutoku:2017wnb}.

Other drivers of neutrino flavor conversion can be more important very close to the merger remnant, in regions where neutrino-matter interactions can still significantly impact the properties of matter outflows. In the presence of high neutrino and/or matter densities, modifications to the Hamiltonian describing the evolution of neutrinos can cause flavor conversions on shorter length scales. The neutrino-matter resonance (NMR)~\cite{2014JPhG...41d4004C,Zhu:2016mwa} occurs when the matter potential and neutrino self-interaction potential intersects, and can be active within a few neutron star radii of a post-merger remnant, though the amount of flavor conversion from the NMR may be insufficient to significantly impact observables from neutron star mergers~\cite{Padilla-Gay:2024wyo}. The fast-flavor instability (FFI)~\cite{PhysRevD.84.053013,Wu:2017drk,Wu:2017qpc,Grohs:2022fyq} is purely due to collective neutrino interactions, i.e. it arises because of corrections to the Hamiltonian for the evolution of neutrinos that depend on the distribution function of other neutrinos. Neutrinos are expected to be unstable to the FFI as soon as a {\it crossing} in the net electron neutrino flux occurs (i.e. at a given point, the flux of neutrinos is larger than the flux of antineutrinos in some directions, but smaller in other directions)~\cite{Izaguirre:2016gsx,Morinaga:2021vmc,Johns:2021taz,Fiorillo:2024bzm}
{\footnote Or rather, instability to the FFI occurs when there is a crossing in the net electron neutrino flux minus the net muon neutrino flux, but the latter is significantly smaller than the former in the merger context, where the distribution functions of $\nu_\mu$ and $\bar\nu_\mu$ are fairly similar before the onset of flavor oscillations. }.
The result of the FFI is typically neutrino flavor conversion that erases such crossings~\cite{Morinaga:2021vmc,Fiorillo:2024qbl}. The FFI has been of particular interest in recent years because neutrinos directly outside of a neutron star remnant and close to the disk can be unstable to the FFI. Specifically, $\nu_e$ and $\bar \nu_e$ decouple from the fluid in different regions, and are produced in different ratios in the disk and remnant neutron star, potentially causing crossings in the net electron neutrino flux outside of the remnant. This is important because flavor conversions close to the remnant are most likely to modify the relative fluxes of $\nu_e$ and $\bar\nu_e$ in ways that would impact absorption in the matter outflows, and thus the neutron-richness of these outflows. Existing global simulations of neutron star mergers at best include very approximate treatments of the FFI~\cite{Li:2021vqj,Fernandez:2022yyv,Just:2022flt}. These results indicate that the FFI may modify predictions for r-process nucleosynthesis, but by how much remains an open question. Finally, the collisional instability~\cite{Johns:2022yqy} is precipitated in denser regions of the remnant by unequal neutrino and antineutrino scattering rates, and may also be active close to the neutrino decoupling regions. It remains unclear whether it can practically impact the outcome of neutron star mergers. In the context of core-collapse supernovae, recent studies have found that the impact of collisional instabilities may be limited~\cite{Akaho:2023brj,Shalgar:2024gjt}.

We refer the reader to the references provided above for more detail on the complex physics involved in neutrino flavor conversions. Reviews of flavor oscillations in mergers are also available in~\cite{Tamborra:2020cul,Richers:2022zug}. In mergers, we are particularly interested in conversions from or into the electron flavor, as an increase or decrease of the flux of $\nu_e$ and $\bar\nu_e$ close to the merger remnant will naturally impact the rate of charged current reactions and thus the composition of the matter, as discussed in the following section.

\subsection{Impact of neutrinos on the evolution of the merger remnant}\label{nu:impact}

In the previous sections, we discussed the important reactions at play within post-merger remnants. Overall, we expect neutrinos to be in equilibrium with the baryonic matter in the center of a post-merger neutron star, then decouple close to the surface -- with $\nu_e$ decoupling closest to the surface and at lower temperature, then $\bar\nu_e$ and $\nu_x$ decoupling at higher densities and temperatures. A post-merger neutron star is then a bright source of neutrino irradiation. In a post-merger accretion disk, neutrinos are more weakly coupled to the matter, with the disk having optical depth to absorption of $\nu_e$ up to $O(1)$, and similar scattering opacities for all species. The hottest parts of the accretion disk, typically shocked tidal arms created immediately after the collision, are the main sources of neutrinos in the disk. Both the remnant neutron star and the hot parts of the accretion disk irradiate the rest of the remnant, as well as any surrounding matter. As neutrinos move away from the remnant, they may also undergo flavor oscillations. With this picture in mind, we can now look at the practical impact of neutrinos on the post-merger remnant.

\subsubsection{Cooling of the remnant}

To first order, the main effect of neutrinos on the post-merger remnant is to cool both the neutron star and the accretion disk. The neutron star remnant has a neutrino luminosity of $10^{52-53}\,{\rm erg/s}$~\cite{Fujibayashi:2020dvr}. On the time scales important for matter ejection from the post-merger remnant (seconds), this is not enough to significantly cool the neutron star, but it provide a continuous source of neutrino irradiation for the rest of the system. Neutrino cooling also becomes important for the evolution of the remnant on longer time scales. Neutrino emission from the shocked regions of the remnant (either in the accretion disk or close to the neutron star surface) can briefly reach $\sim 10^{54}\,{\rm erg/s}$, but this emission lasts only $O(10\,{\rm ms})$~\cite{Deaton:2013sla,Sekiguchi:2016bjd,Fujibayashi:2020dvr,Radice:2021jtw}. 

Emission from the accretion disk may remain as high as $10^{52-53}\,{\rm erg/s}$ for $O(100\,{\rm ms})$, the timescale for viscous spreading of the disk~\cite{Fernandez:2020oow,Shibata:2021xmo,Fujibayashi:2022ftg}. To assess the impact of neutrino cooling, it is good to compare that energy to the rate at which the accretion disk gains energy due to the increase of the binding energy of particles as they approach the compact object, i.e. $\sim G\dot M M_{\rm rem}/R$ (with $\dot M$ the accretion rate, $M_{\rm rem}$ the mass of the remnant, and $R$ the radius of the inner edge of the disk). At this luminosity and for typical accretion rates in post-merger remnants (a few percents of a solar mass per second or more in the first $0.1$ second~\cite{Fernandez:2018kax}), the change in the binding energy of the matter as it accretes on the central object is slightly larger than the energy lost to neutrinos, so that part of that energy serves to heat the accretion disk. An efficiently cooled accretion disk would become very thin (height over radius ratio $H/r\ll 1$), due to the lack of vertical pressure support. This is seen for example in thin disks around supermassive black holes, which can be efficiently cooled by photon emission. In post-merger remnants, neutrino cooling does help to reduce the height of the disk, but only down to $H/r\sim 0.2$. Neutrino cooling has an important effect on the evolution of the disk, but is not sufficient to efficiently radiate the energy generated by the accretion process. After $\sim 0.1\,{\rm s}$, as the accretion rate decreases, neutrino cooling becomes less efficient. Indeed, the main cooling reactions $n+e^+ \rightarrow p+\bar\nu_e$ and $p+e^- \rightarrow n + \nu_e$ have luminosities scaling as $(n_n n_{e^+})$ and $(n_p n_{e^-})$, respectively, while the thermal energy of the disk approximately scales linearly with its density. As soon as the disk density decreases, neutrino cooling becomes inefficient and the accretion disk puffs up to $H/r\sim 1$~\cite{Just:2021cls,De:2020jdt}.
Viscous heating and $\alpha$-particle formation release energy, with no cooling mechanism to compensate. Over seconds timescale, this leads to the ejection of a significant fraction of the disk mass ($\sim 5\%-20\%$)~\cite{Fernandez:2013tya}.

Overall, we thus have two main phases to the evolution of the post-merger remnant, with neutrino emission being crucial to the transition between the two. First, we have a compact object surrounded by a {\it neutrino-driven accretion flow} (NDAF). In an NDAF, neutrino cooling is sufficient to partially reduce the thickness of the disk, and both the disk and the neutron star remnant (if present) contribute significantly to the neutrino luminosity. On the accretion time scale $O(0.1\,{\rm s})$, the mass accretion rate and density of the disk decrease. Neutrino cooling from the disk becomes inefficient and we have an advection dominated accretion flow (ADAF). Neutrino emission then comes nearly entirely from the neutron star remnant, if the remnant has not collapsed to a black hole. The exact time of the transition depends on the mass and history of the accretion disk. For a disk around a black hole of mass $M_{\rm BH}$ it is expected to occur when the mass accretion rate drops below the ``ignition threshold''~\cite{Metzger:2007kj}
\begin{equation}
\dot M_{\rm ign} = 0.002 M_\odot s^{-1} \left(\frac{M_{\rm BH}}{3M_\odot}\right)^{4/3}
\left(\frac{\alpha_{\rm vis}}{0.02}\right)^{5/3}
\end{equation}
with $\alpha_{\rm vis}$ a parameter measuring the strength of viscous heating in the disk. 
The vertical structure of the disk has important consequences for the ability of the remnant disk to launch winds and eject matter, its accretion rate, or even the speed at which it aligns with the equatorial plane of the central remnant in cases where the disk is initially misaligned (i.e. if one of the original compact object had a rotation axis misaligned with the orbital angular momentum of the binary). Neutrinos thus play a crucial role in the dynamics and thermodynamics of the post-merger remnant for as long as we are in the NDAF regime.

\subsubsection{Changes in the composition of the fluid}

\begin{figure}[h]
	\centering
	\includegraphics[width=0.9\linewidth]{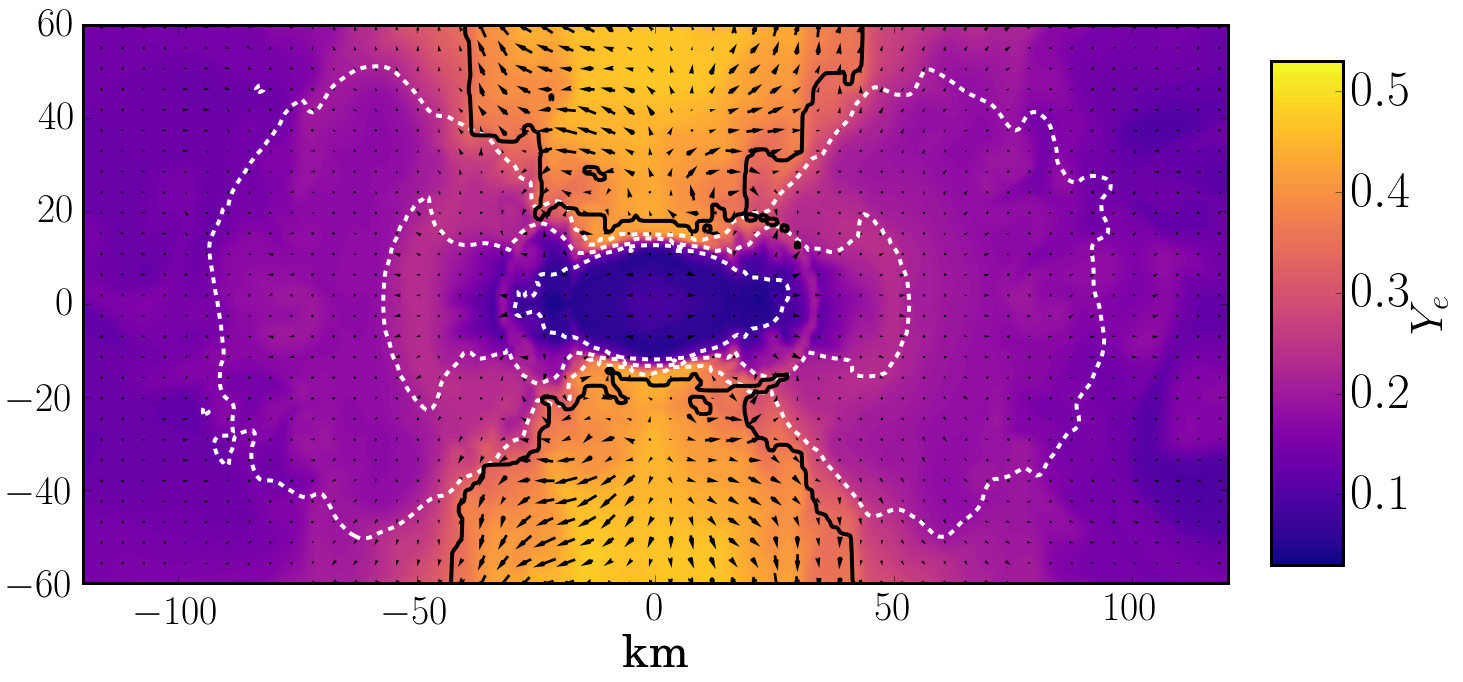}
	\caption{Vertical slice through a post-merger neutron star remnant $10\,{\rm ms}$ after merger (from the same simulation as Fig.~\ref{fig:neutrinosphere}). The dotted white lines show density contours of $10^{11,12,13}\,{\rm g/cm^3}$; the solid black line marks the boundary between bound and ejected matter. Arrows show the in-plane components of the velocity of the fluid, and the color scale shows the composition of the remnant. We see the highly neutron-rich core at the center, the moderately neutron-rich disk in the equatorial plane, and higher $Y_e$ outflows in the polar regions. Figure reproduced from~\cite{Foucart:2016rxm}}
	\label{fig:ye}
\end{figure}

The second important effect of neutrinos in post-merger remnants, and the one with the most direct impact on observables, is to drive changes in the composition of the post-merger remnant, i.e. its neutron-richness. Before merger, most of the matter has $Y_e\lesssim 0.1$. As soon as the density of the matter decreases, or the temperature increases, it becomes energetically favorable for $Y_e$ to increase. Charged current reactions transforming neutrons into protons are favored over their inverse reactions. In the cold matter ejected during the tidal disruption of a neutron star, neutrino emission and absorption are rare. Accordingly, $Y_e$ does not change much, and the matter remains very neutron rich. In the accretion disk and in the hot outflows ejected from the disk and neutron star remnant, on the other hand, neutrino-matter interactions quickly lead to an increase in $Y_e$~\cite{Wanajo:2014}. Fig.~\ref{fig:ye} illustrates the large difference in composition between a remnant neutron star (very neutron rich), an accretion disk (less neutron rich), and hot matter ouflows ($Y_e>0.3$).

In the accretion disk, charged current reactions are initially fast enough for the matter to reach its equilibrium composition of $Y_e \sim (0.1-0.2)$ within a few milliseconds. At later times, the composition of the disk evolves because the equilibrium composition varies with the density and temperature of the matter, increasing with decreasing density. After $O(0.1\,{\rm s})$, weak interactions become too slow to maintain the equilibrium electron fraction, and the $Y_e$ of the disk freezes out. The expected value of $Y_e$ at freeze-out is $\sim 0.3$~\cite{Just:2021cls,De:2020jdt}. This is a particularly important process when considering r-process nucleosynthesis and kilonova emission. As mentioned earlier, $\sim 5\%-20\%$ of the initial accretion disk is unbound during the ADAF phase, and that matter will be ejected with an electron fraction close to the freeze-out value.

Neutrino emissions and absorptions also have a strong impact on the asymptotic electron fraction of hot matter ejected from the surface of the neutron star or the inner regions of the accretion disk (largely through magnetically-driven winds), as well as from the collision of the two neutron stars. For matter irradiated by $\nu_e$ and $\bar\nu_e$, the equilibrium electron fraction is~\cite{1993PhRvL..71.1965Q}
\begin{equation}
Y_e^{\rm eq} = \frac{1}{1+\frac{L_{\bar\nu_e}\epsilon_{\bar\nu_e}}{L_{\nu_e}\epsilon_{\nu_e}}}
\end{equation}
with $L_{\nu}$ and $\epsilon_\nu$ the luminosity and average energy of neutrinos.
In post-merger remnants, the electron antineutrinos are typically more energetic than the neutrinos, and $Y_e^{\rm eq} \sim (0.3-0.5)$. One might expect the electron fraction of these outflows to be large enough to prevent the formation of the higher mass r-process elements. However, there is a competing effect at play: the ejected matter is relativistic ($v\sim 0.1c-0.3c$ for the bulk of the ejecta). As it expands, it rapidly cools down and decreases in density while being subjected to lower neutrino irradiation. This reduces the efficiency of weak interactions. Fast moving ejecta may not have enough time to evolve to its equilibrium electron fraction. In practice, merger simulations show that the hot matter ejected during the collision of the two neutron stars has a broad range of $Y_e~\sim (0.2-0.5)$~\cite{Wanajo:2014,Sekiguchi:2016bjd,Foucart:2020qjb,Radice:2021jtw}. Matter ejected through magnetically-driven winds has more uncertain properties, with higher electron fractions reached in the presence of a hot neutron star, while matter ejected in black hole-disk systems may produce a broad range of $Y_e$ with peak composition closer to the lower end of that $Y_e$ range~\cite{Miller:2019dpt,Hayashi:2021oxy}. Considering that r-process yields are very sensitive to variations of $Y_e$ in the $0.2-0.3$ range, this remains a significant source of uncertainty when trying to predict nucleosynthetic yields from neutron star mergers. These uncertainties are compounded by the unknown impact of neutrino flavor oscillations: flavor oscillations could change the ratio $\frac{L_{\bar\nu_e}\epsilon_{\bar\nu_e}}{L_{\nu_e}\epsilon_{\nu_e}}$, and thus $Y_e^{\rm eq}$.

\subsubsection{Energy deposition from pair annihilation}

The last impact of neutrinos on the post-merger remnant that we will discuss here is the deposition of energy by $\nu\bar\nu$ pair annihilation in low-density regions above the remnant. Pair annihilation occurs everywhere, but in high-density regions it generally has less of an impact on the evolution of the fluid than charged-current reactions. In low-density regions, if the flux of $\nu$ and $\bar\nu$ is high, this is no longer the case. The efficiency of the conversion of neutrinos into $e^+e^-$ pairs depends a lot on the geometry of the system. As mentioned in Sec.~\ref{sec:pairs}, the annihilation rate is proportional to $(1-\cos\theta)^2$, with $\theta$ the angle between the direction of propagation of the neutrinos. Around the rotation axis of the remnant, we have neutrinos coming from different regions of the accretion disk and from the hot neutron star, and relatively low matter densities. This is thus where pair annihilation will be the most important. Estimates from numerical simulations indicate that $O(1\%)$ of the energy emitted in neutrinos by the accretion disk can be converted into electron-positron pairs in the polar regions~\cite{Just:2015dba}. We would thus expect at most $10^{49-50}\,{\rm ergs}$ deposited in the polar regions. Neutrino-antineutrino pair annihilation was initially proposed as a mechanism for the production of short gamma-ray bursts~\cite{Eichler:1989ve}; yet only the lower-energy bursts can be explained by this mechanism given the predicted energy deposition. Pair annihilation does however have a role in accelerating the matter within the polar regions to relativistic speeds ($v\sim 0.9c$)~\cite{Fujibayashi:2017xsz}. This acceleration could help clear the polar regions of baryons, thus facilitating the production of a relativistic jet and gamma-ray burst.

\section{Impact on observables}\label{obs}

Let us now briefly return to the observables from neutron star mergers, and review how neutrinos may impact these observations, as well as our understanding of important open questions in nuclear astrophysics.

The matter ejected by neutron star mergers ranges from $Y_e<0.1$ (cold ejecta from tidal disruption), to $Y_e\sim (0.2-0.3)$ (outflows in ADAF regime, magnetically-driven winds from black hole-disk systems), to $Y_e>0.3$ (outflows around hot neutron stars). The exact electron fraction reached is mainly determined by neutrino-matter interactions. Crucially, the electron fraction of the outflows determines the outcome of r-process nucleosynthesis. The lower $Y_e$ outflows will produce an overabundance of heavy r-process elements, including lanthanides and actinides. The higher $Y_e$ outflows produce lower mass r-process elements. The outflows in the intermediate range have uncertain nucleosynthetic products, that current models of neutron star mergers cannot reliably predict. All three types of outflows likely exist in neutron star mergers.

Variations in the electron fraction of the outflows also lead to very different kilonova signals for the different types of outflows. More massive outflows lead to brighter and longer duration kilonovae; more neutron-rich outflows lead to redder, dimmer, and longer duration signals. Finally, the velocity of the outflows is also correlated with the timescale and brightness of the kilonova emission~\cite{Barnes:2013wka}. Observations of kilonovae, if properly understood, can thus tell us about both r-process nucleosynthesis in the outflows and the properties of the outflows themselves. As discussed earlier, this can provide useful information about the properties of the merging compact objects and of the nuclear matter around and above nuclear saturation density.

Finally, we have seen that $\nu\bar\nu$ annihilations can deposit energy in low-density polar regions above a neutron star remnant. This energy deposition drives matter to relativistic velocities, and may help in the production of relativistic jets -- though it is likely not the dominant process behind the production of the high Lorentz factor outflows powering short gamma-ray bursts.

\section{Conclusions}
\label{sec:conclusions}

Neutron star mergers are highly energetic events powering gravitational wave and electromagnetic signals observable from hundreds of millions to billions of light-years away. They may play an important role as the production site of many heavy atomic nuclei in the Universe, and allow us to study the properties of dense cold matter in conditions that cannot be reproduced on Earth. Neutrinos play a crucial role in these systems. Neutrino-matter interactions are the main determinant of the neutron-richness of the matter ejected by neutron star mergers, which itself will determine the elements produced in that ejecta -- elements that will then propagate through the surrounding galaxy, and eventually become part of newly formed stars and planets. Neutrino-matter interactions also determine the properties of the electromagnetic signal powered by radioactive decays of these newly formed elements.

Beyond these important questions in nuclear astrophysics, neutron star mergers also provide us with a rare environment in which neutrinos can be tightly coupled to other particles, and reach high-enough densities that collective neutrino effects might be observable. This makes neutrino effects in neutron star mergers harder to model: because neutrinos are in statistical equilibrium with other particles in the densest regions of neutron stars, partially coupled to matter in accretion disks, and free-streaming elsewhere, they can only be studied through the evolution of Boltzmann's equations of radiation transport. This remains a very expensive problem for modern merger simulations, and the development of improved neutrino transport algorithm for the evolution of neutron star mergers and post-merger remnants is a very active field of research today~\cite{Wanajo:2014,Foucart:2016rxm,Miller:2019dpt,Radice:2021jtw,Foucart:2022kon,Kawaguchi:2022tae,Izquierdo:2023fub}. Neutrino flavor transformations through collective oscillations triggered by small scale instabilities adds another source of uncertainty in our predictions for neutrino effects in mergers, but also a fascinating window into the behavior of neutrinos at densities that can only be naturally reached in neutron star mergers, supernovae, and the early Universe.

The last decade has seen a rapid increase in our understanding of neutrino effects in mergers, thanks in part to theoretical work and numerical simulations, but also through the first observations of neutron star mergers and a rapidly growing number of observations of r-process abundancies in diverse environments. Nevertheless, a number of important open questions remain. Despite evidence that some r-process elements are produced in neutron star mergers, we remain unsure of whether mergers are (one of) the main source(s) of r-process elements in the Universe or not. Additionally, our understanding of neutron star mergers in general and of neutrino-matter interactions in particular remains an important limitation to our ability to use these mergers to extract information about the properties of dense nuclear matter. Improving on the state of the art will require a combination of theoretical and computational work as well as higher quality and more common observations. The expected sensitivity increases of current gravitational wave detectors and construction of third generation detectors~\cite{Maggiore:2019uih,Evans:2023euw} should help with our understanding of dense matter, while the study of other open questions in nuclear physics associated with mergers likely require electromagnetic observations and thus rapid follow-up of gravitational wave events with electromagnetic telescopes.

\begin{acknowledgments}
 We thank Rodrigo Fernandez, Irene Tamborra, Somdutta Ghosh, Christian Fischer, and Damiano Forillo for discussions and comments provided on drafts of this chapter. This work was performed in part at the Aspen Center for Physics, which is supported by National Science Foundation grant PHY-2210452. FF gratefully acknowledges support from the Department of Energy, Office of Science, Office of Nuclear Physics, under contract number DE-AC02-05CH11231 and DE-SC0025023, from NASA through grant 80NSSC22K0719, and from the NSF through grant AST-2107932.
\end{acknowledgments}

\bibliography{reference}

\begin{thebibliography}{88}%
\makeatletter
\providecommand \@ifxundefined [1]{%
 \@ifx{#1\undefined}
}%
\providecommand \@ifnum [1]{%
 \ifnum #1\expandafter \@firstoftwo
 \else \expandafter \@secondoftwo
 \fi
}%
\providecommand \@ifx [1]{%
 \ifx #1\expandafter \@firstoftwo
 \else \expandafter \@secondoftwo
 \fi
}%
\providecommand \natexlab [1]{#1}%
\providecommand \enquote  [1]{``#1''}%
\providecommand \bibnamefont  [1]{#1}%
\providecommand \bibfnamefont [1]{#1}%
\providecommand \citenamefont [1]{#1}%
\providecommand \href@noop [0]{\@secondoftwo}%
\providecommand \href [0]{\begingroup \@sanitize@url \@href}%
\providecommand \@href[1]{\@@startlink{#1}\@@href}%
\providecommand \@@href[1]{\endgroup#1\@@endlink}%
\providecommand \@sanitize@url [0]{\catcode `\\12\catcode `\$12\catcode
  `\&12\catcode `\#12\catcode `\^12\catcode `\_12\catcode `\%12\relax}%
\providecommand \@@startlink[1]{}%
\providecommand \@@endlink[0]{}%
\providecommand \url  [0]{\begingroup\@sanitize@url \@url }%
\providecommand \@url [1]{\endgroup\@href {#1}{\urlprefix }}%
\providecommand \urlprefix  [0]{URL }%
\providecommand \Eprint [0]{\href }%
\providecommand \doibase [0]{https://doi.org/}%
\providecommand \selectlanguage [0]{\@gobble}%
\providecommand \bibinfo  [0]{\@secondoftwo}%
\providecommand \bibfield  [0]{\@secondoftwo}%
\providecommand \translation [1]{[#1]}%
\providecommand \BibitemOpen [0]{}%
\providecommand \bibitemStop [0]{}%
\providecommand \bibitemNoStop [0]{.\EOS\space}%
\providecommand \EOS [0]{\spacefactor3000\relax}%
\providecommand \BibitemShut  [1]{\csname bibitem#1\endcsname}%
\let\auto@bib@innerbib\@empty
\bibitem [{\citenamefont {Lattimer}(2021)}]{Lattimer:2021emm}%
  \BibitemOpen
  \bibfield  {author} {\bibinfo {author} {\bibfnamefont {J.~M.}\ \bibnamefont
  {Lattimer}},\ }\bibfield  {title} {\bibinfo {title} {{Neutron Stars and the
  Nuclear Matter Equation of State}},\ }\href
  {https://doi.org/10.1146/annurev-nucl-102419-124827} {\bibfield  {journal}
  {\bibinfo  {journal} {Ann. Rev. Nucl. Part. Sci.}\ }\textbf {\bibinfo
  {volume} {71}},\ \bibinfo {pages} {433} (\bibinfo {year} {2021})}\BibitemShut
  {NoStop}%
\bibitem [{\citenamefont {Bardeen}\ \emph {et~al.}(1972)\citenamefont
  {Bardeen}, \citenamefont {Press},\ and\ \citenamefont
  {Teukolsky}}]{Bardeen:1972fi}%
  \BibitemOpen
  \bibfield  {author} {\bibinfo {author} {\bibfnamefont {J.~M.}\ \bibnamefont
  {Bardeen}}, \bibinfo {author} {\bibfnamefont {W.~H.}\ \bibnamefont {Press}},\
  and\ \bibinfo {author} {\bibfnamefont {S.~A.}\ \bibnamefont {Teukolsky}},\
  }\bibfield  {title} {\bibinfo {title} {{Rotating black holes: Locally
  nonrotating frames, energy extraction, and scalar synchrotron radiation}},\
  }\href {https://doi.org/10.1086/151796} {\bibfield  {journal} {\bibinfo
  {journal} {Astrophys. J.}\ }\textbf {\bibinfo {volume} {178}},\ \bibinfo
  {pages} {347} (\bibinfo {year} {1972})}\BibitemShut {NoStop}%
\bibitem [{\citenamefont {{{\"O}zel}}\ \emph {et~al.}(2010)\citenamefont
  {{{\"O}zel}}, \citenamefont {{Psaltis}}, \citenamefont {{Narayan}},\ and\
  \citenamefont {{McClintock}}}]{2010ApJ...725.1918O}%
  \BibitemOpen
  \bibfield  {author} {\bibinfo {author} {\bibfnamefont {F.}~\bibnamefont
  {{{\"O}zel}}}, \bibinfo {author} {\bibfnamefont {D.}~\bibnamefont
  {{Psaltis}}}, \bibinfo {author} {\bibfnamefont {R.}~\bibnamefont
  {{Narayan}}},\ and\ \bibinfo {author} {\bibfnamefont {J.~E.}\ \bibnamefont
  {{McClintock}}},\ }\bibfield  {title} {\bibinfo {title} {{The Black Hole Mass
  Distribution in the Galaxy}},\ }\href
  {https://doi.org/10.1088/0004-637X/725/2/1918} {\bibfield  {journal}
  {\bibinfo  {journal} {Astrophys.J.}\ }\textbf {\bibinfo {volume} {725}},\
  \bibinfo {pages} {1918} (\bibinfo {year} {2010})},\ \Eprint
  {https://arxiv.org/abs/1006.2834} {arXiv:1006.2834 [astro-ph.GA]}
  \BibitemShut {NoStop}%
\bibitem [{\citenamefont {Farr}\ \emph {et~al.}(2011)\citenamefont {Farr},
  \citenamefont {Sravan}, \citenamefont {Cantrell}, \citenamefont {Kreidberg},
  \citenamefont {Bailyn}, \citenamefont {Mandel},\ and\ \citenamefont
  {Kalogera}}]{Farr:2010tu}%
  \BibitemOpen
  \bibfield  {author} {\bibinfo {author} {\bibfnamefont {W.~M.}\ \bibnamefont
  {Farr}}, \bibinfo {author} {\bibfnamefont {N.}~\bibnamefont {Sravan}},
  \bibinfo {author} {\bibfnamefont {A.}~\bibnamefont {Cantrell}}, \bibinfo
  {author} {\bibfnamefont {L.}~\bibnamefont {Kreidberg}}, \bibinfo {author}
  {\bibfnamefont {C.~D.}\ \bibnamefont {Bailyn}}, \bibinfo {author}
  {\bibfnamefont {I.}~\bibnamefont {Mandel}},\ and\ \bibinfo {author}
  {\bibfnamefont {V.}~\bibnamefont {Kalogera}},\ }\bibfield  {title} {\bibinfo
  {title} {{The Mass Distribution of Stellar-Mass Black Holes}},\ }\href
  {https://doi.org/10.1088/0004-637X/741/2/103} {\bibfield  {journal} {\bibinfo
   {journal} {Astrophys. J.}\ }\textbf {\bibinfo {volume} {741}},\ \bibinfo
  {pages} {103} (\bibinfo {year} {2011})},\ \Eprint
  {https://arxiv.org/abs/1011.1459} {arXiv:1011.1459 [astro-ph.GA]}
  \BibitemShut {NoStop}%
\bibitem [{\citenamefont {Abbott}\ \emph {et~al.}(2023)\citenamefont {Abbott}
  \emph {et~al.}}]{KAGRA:2021vkt}%
  \BibitemOpen
  \bibfield  {author} {\bibinfo {author} {\bibfnamefont {R.}~\bibnamefont
  {Abbott}} \emph {et~al.} (\bibinfo {collaboration} {KAGRA, VIRGO, LIGO
  Scientific}),\ }\bibfield  {title} {\bibinfo {title} {{GWTC-3: Compact Binary
  Coalescences Observed by LIGO and Virgo during the Second Part of the Third
  Observing Run}},\ }\href {https://doi.org/10.1103/PhysRevX.13.041039}
  {\bibfield  {journal} {\bibinfo  {journal} {Phys. Rev. X}\ }\textbf {\bibinfo
  {volume} {13}},\ \bibinfo {pages} {041039} (\bibinfo {year} {2023})},\
  \Eprint {https://arxiv.org/abs/2111.03606} {arXiv:2111.03606 [gr-qc]}
  \BibitemShut {NoStop}%
\bibitem [{\citenamefont {Abbott}\ \emph {et~al.}(2016)\citenamefont {Abbott}
  \emph {et~al.}}]{LIGOScientific:2016aoc}%
  \BibitemOpen
  \bibfield  {author} {\bibinfo {author} {\bibfnamefont {B.~P.}\ \bibnamefont
  {Abbott}} \emph {et~al.} (\bibinfo {collaboration} {LIGO Scientific,
  Virgo}),\ }\bibfield  {title} {\bibinfo {title} {{Observation of
  Gravitational Waves from a Binary Black Hole Merger}},\ }\href
  {https://doi.org/10.1103/PhysRevLett.116.061102} {\bibfield  {journal}
  {\bibinfo  {journal} {Phys. Rev. Lett.}\ }\textbf {\bibinfo {volume} {116}},\
  \bibinfo {pages} {061102} (\bibinfo {year} {2016})},\ \Eprint
  {https://arxiv.org/abs/1602.03837} {arXiv:1602.03837 [gr-qc]} \BibitemShut
  {NoStop}%
\bibitem [{\citenamefont {Abbott}\ \emph {et~al.}(2017)\citenamefont {Abbott}
  \emph {et~al.}}]{LIGOScientific:2017vwq}%
  \BibitemOpen
  \bibfield  {author} {\bibinfo {author} {\bibfnamefont {B.~P.}\ \bibnamefont
  {Abbott}} \emph {et~al.} (\bibinfo {collaboration} {LIGO Scientific,
  Virgo}),\ }\bibfield  {title} {\bibinfo {title} {{GW170817: Observation of
  Gravitational Waves from a Binary Neutron Star Inspiral}},\ }\href
  {https://doi.org/10.1103/PhysRevLett.119.161101} {\bibfield  {journal}
  {\bibinfo  {journal} {Phys. Rev. Lett.}\ }\textbf {\bibinfo {volume} {119}},\
  \bibinfo {pages} {161101} (\bibinfo {year} {2017})},\ \Eprint
  {https://arxiv.org/abs/1710.05832} {arXiv:1710.05832 [gr-qc]} \BibitemShut
  {NoStop}%
\bibitem [{\citenamefont {Abbott}\ \emph {et~al.}(2021)\citenamefont {Abbott}
  \emph {et~al.}}]{LIGOScientific:2021qlt}%
  \BibitemOpen
  \bibfield  {author} {\bibinfo {author} {\bibfnamefont {R.}~\bibnamefont
  {Abbott}} \emph {et~al.} (\bibinfo {collaboration} {LIGO Scientific, KAGRA,
  VIRGO}),\ }\bibfield  {title} {\bibinfo {title} {{Observation of
  Gravitational Waves from Two Neutron Star\textendash{}Black Hole
  Coalescences}},\ }\href {https://doi.org/10.3847/2041-8213/ac082e} {\bibfield
   {journal} {\bibinfo  {journal} {Astrophys. J. Lett.}\ }\textbf {\bibinfo
  {volume} {915}},\ \bibinfo {pages} {L5} (\bibinfo {year} {2021})},\ \Eprint
  {https://arxiv.org/abs/2106.15163} {arXiv:2106.15163 [astro-ph.HE]}
  \BibitemShut {NoStop}%
\bibitem [{\citenamefont {Foucart}\ \emph {et~al.}(2023)\citenamefont
  {Foucart}, \citenamefont {Duez}, \citenamefont {Haas}, \citenamefont
  {Kidder}, \citenamefont {Pfeiffer}, \citenamefont {Scheel},\ and\
  \citenamefont {Spira-Savett}}]{Foucart:2022kon}%
  \BibitemOpen
  \bibfield  {author} {\bibinfo {author} {\bibfnamefont {F.}~\bibnamefont
  {Foucart}}, \bibinfo {author} {\bibfnamefont {M.~D.}\ \bibnamefont {Duez}},
  \bibinfo {author} {\bibfnamefont {R.}~\bibnamefont {Haas}}, \bibinfo {author}
  {\bibfnamefont {L.~E.}\ \bibnamefont {Kidder}}, \bibinfo {author}
  {\bibfnamefont {H.~P.}\ \bibnamefont {Pfeiffer}}, \bibinfo {author}
  {\bibfnamefont {M.~A.}\ \bibnamefont {Scheel}},\ and\ \bibinfo {author}
  {\bibfnamefont {E.}~\bibnamefont {Spira-Savett}},\ }\bibfield  {title}
  {\bibinfo {title} {{General relativistic simulations of collapsing binary
  neutron star mergers with Monte~Carlo neutrino transport}},\ }\href
  {https://doi.org/10.1103/PhysRevD.107.103055} {\bibfield  {journal} {\bibinfo
   {journal} {Phys. Rev. D}\ }\textbf {\bibinfo {volume} {107}},\ \bibinfo
  {pages} {103055} (\bibinfo {year} {2023})},\ \Eprint
  {https://arxiv.org/abs/2210.05670} {arXiv:2210.05670 [astro-ph.HE]}
  \BibitemShut {NoStop}%
\bibitem [{\citenamefont {Kasen}\ \emph {et~al.}(2017)\citenamefont {Kasen},
  \citenamefont {Metzger}, \citenamefont {Barnes}, \citenamefont {Quataert},\
  and\ \citenamefont {Ramirez-Ruiz}}]{Kasen:2017sxr}%
  \BibitemOpen
  \bibfield  {author} {\bibinfo {author} {\bibfnamefont {D.}~\bibnamefont
  {Kasen}}, \bibinfo {author} {\bibfnamefont {B.}~\bibnamefont {Metzger}},
  \bibinfo {author} {\bibfnamefont {J.}~\bibnamefont {Barnes}}, \bibinfo
  {author} {\bibfnamefont {E.}~\bibnamefont {Quataert}},\ and\ \bibinfo
  {author} {\bibfnamefont {E.}~\bibnamefont {Ramirez-Ruiz}},\ }\bibfield
  {title} {\bibinfo {title} {{Origin of the heavy elements in binary
  neutron-star mergers from a gravitational wave event}},\ }\href
  {https://doi.org/10.1038/nature24453} {\bibfield  {journal} {\bibinfo
  {journal} {Nature}\ }\textbf {\bibinfo {volume} {551}},\ \bibinfo {pages}
  {80} (\bibinfo {year} {2017})},\ \Eprint {https://arxiv.org/abs/1710.05463}
  {arXiv:1710.05463 [astro-ph.HE]} \BibitemShut {NoStop}%
\bibitem [{\citenamefont {{Pian}}\ \emph {et~al.}(2017)\citenamefont {{Pian}},
  \citenamefont {{D'Avanzo}}, \citenamefont {{Benetti}}, \citenamefont
  {{Branchesi}}, \citenamefont {{Brocato}}, \citenamefont {{Campana}},
  \citenamefont {{Cappellaro}}, \citenamefont {{Covino}}, \citenamefont
  {{D'Elia}}, \citenamefont {{Fynbo}}, \citenamefont {{Getman}}, \citenamefont
  {{Ghirlanda}}, \citenamefont {{Ghisellini}}, \citenamefont {{Grado}},
  \citenamefont {{Greco}}, \citenamefont {{Hjorth}}, \citenamefont
  {{Kouveliotou}}, \citenamefont {{Levan}}, \citenamefont {{Limatola}},
  \citenamefont {{Malesani}}, \citenamefont {{Mazzali}}, \citenamefont
  {{Melandri}}, \citenamefont {{M{\o}ller}}, \citenamefont {{Nicastro}},
  \citenamefont {{Palazzi}}, \citenamefont {{Piranomonte}}, \citenamefont
  {{Rossi}}, \citenamefont {{Salafia}}, \citenamefont {{Selsing}},
  \citenamefont {{Stratta}}, \citenamefont {{Tanaka}}, \citenamefont
  {{Tanvir}}, \citenamefont {{Tomasella}}, \citenamefont {{Watson}},
  \citenamefont {{Yang}}, \citenamefont {{Amati}}, \citenamefont {{Antonelli}},
  \citenamefont {{Ascenzi}}, \citenamefont {{Bernardini}}, \citenamefont
  {{Bo{\"e}r}}, \citenamefont {{Bufano}}, \citenamefont {{Bulgarelli}},
  \citenamefont {{Capaccioli}}, \citenamefont {{Casella}}, \citenamefont
  {{Castro-Tirado}}, \citenamefont {{Chassande-Mottin}}, \citenamefont
  {{Ciolfi}}, \citenamefont {{Copperwheat}}, \citenamefont {{Dadina}},
  \citenamefont {{De Cesare}}, \citenamefont {{di Paola}}, \citenamefont
  {{Fan}}, \citenamefont {{Gendre}}, \citenamefont {{Giuffrida}}, \citenamefont
  {{Giunta}}, \citenamefont {{Hunt}}, \citenamefont {{Israel}}, \citenamefont
  {{Jin}}, \citenamefont {{Kasliwal}}, \citenamefont {{Klose}}, \citenamefont
  {{Lisi}}, \citenamefont {{Longo}}, \citenamefont {{Maiorano}}, \citenamefont
  {{Mapelli}}, \citenamefont {{Masetti}}, \citenamefont {{Nava}}, \citenamefont
  {{Patricelli}}, \citenamefont {{Perley}}, \citenamefont {{Pescalli}},
  \citenamefont {{Piran}}, \citenamefont {{Possenti}}, \citenamefont
  {{Pulone}}, \citenamefont {{Razzano}}, \citenamefont {{Salvaterra}},
  \citenamefont {{Schipani}}, \citenamefont {{Spera}}, \citenamefont
  {{Stamerra}}, \citenamefont {{Stella}}, \citenamefont {{Tagliaferri}},
  \citenamefont {{Testa}}, \citenamefont {{Troja}}, \citenamefont {{Turatto}},
  \citenamefont {{Vergani}},\ and\ \citenamefont
  {{Vergani}}}]{2017Natur.551...67P}%
  \BibitemOpen
  \bibfield  {author} {\bibinfo {author} {\bibfnamefont {E.}~\bibnamefont
  {{Pian}}}, \bibinfo {author} {\bibfnamefont {P.}~\bibnamefont {{D'Avanzo}}},
  \bibinfo {author} {\bibfnamefont {S.}~\bibnamefont {{Benetti}}}, \bibinfo
  {author} {\bibfnamefont {M.}~\bibnamefont {{Branchesi}}}, \bibinfo {author}
  {\bibfnamefont {E.}~\bibnamefont {{Brocato}}}, \bibinfo {author}
  {\bibfnamefont {S.}~\bibnamefont {{Campana}}}, \bibinfo {author}
  {\bibfnamefont {E.}~\bibnamefont {{Cappellaro}}}, \bibinfo {author}
  {\bibfnamefont {S.}~\bibnamefont {{Covino}}}, \bibinfo {author}
  {\bibfnamefont {V.}~\bibnamefont {{D'Elia}}}, \bibinfo {author}
  {\bibfnamefont {J.~P.~U.}\ \bibnamefont {{Fynbo}}}, \bibinfo {author}
  {\bibfnamefont {F.}~\bibnamefont {{Getman}}}, \bibinfo {author}
  {\bibfnamefont {G.}~\bibnamefont {{Ghirlanda}}}, \bibinfo {author}
  {\bibfnamefont {G.}~\bibnamefont {{Ghisellini}}}, \bibinfo {author}
  {\bibfnamefont {A.}~\bibnamefont {{Grado}}}, \bibinfo {author} {\bibfnamefont
  {G.}~\bibnamefont {{Greco}}}, \bibinfo {author} {\bibfnamefont
  {J.}~\bibnamefont {{Hjorth}}}, \bibinfo {author} {\bibfnamefont
  {C.}~\bibnamefont {{Kouveliotou}}}, \bibinfo {author} {\bibfnamefont
  {A.}~\bibnamefont {{Levan}}}, \bibinfo {author} {\bibfnamefont
  {L.}~\bibnamefont {{Limatola}}}, \bibinfo {author} {\bibfnamefont
  {D.}~\bibnamefont {{Malesani}}}, \bibinfo {author} {\bibfnamefont {P.~A.}\
  \bibnamefont {{Mazzali}}}, \bibinfo {author} {\bibfnamefont {A.}~\bibnamefont
  {{Melandri}}}, \bibinfo {author} {\bibfnamefont {P.}~\bibnamefont
  {{M{\o}ller}}}, \bibinfo {author} {\bibfnamefont {L.}~\bibnamefont
  {{Nicastro}}}, \bibinfo {author} {\bibfnamefont {E.}~\bibnamefont
  {{Palazzi}}}, \bibinfo {author} {\bibfnamefont {S.}~\bibnamefont
  {{Piranomonte}}}, \bibinfo {author} {\bibfnamefont {A.}~\bibnamefont
  {{Rossi}}}, \bibinfo {author} {\bibfnamefont {O.~S.}\ \bibnamefont
  {{Salafia}}}, \bibinfo {author} {\bibfnamefont {J.}~\bibnamefont
  {{Selsing}}}, \bibinfo {author} {\bibfnamefont {G.}~\bibnamefont
  {{Stratta}}}, \bibinfo {author} {\bibfnamefont {M.}~\bibnamefont {{Tanaka}}},
  \bibinfo {author} {\bibfnamefont {N.~R.}\ \bibnamefont {{Tanvir}}}, \bibinfo
  {author} {\bibfnamefont {L.}~\bibnamefont {{Tomasella}}}, \bibinfo {author}
  {\bibfnamefont {D.}~\bibnamefont {{Watson}}}, \bibinfo {author}
  {\bibfnamefont {S.}~\bibnamefont {{Yang}}}, \bibinfo {author} {\bibfnamefont
  {L.}~\bibnamefont {{Amati}}}, \bibinfo {author} {\bibfnamefont {L.~A.}\
  \bibnamefont {{Antonelli}}}, \bibinfo {author} {\bibfnamefont
  {S.}~\bibnamefont {{Ascenzi}}}, \bibinfo {author} {\bibfnamefont {M.~G.}\
  \bibnamefont {{Bernardini}}}, \bibinfo {author} {\bibfnamefont
  {M.}~\bibnamefont {{Bo{\"e}r}}}, \bibinfo {author} {\bibfnamefont
  {F.}~\bibnamefont {{Bufano}}}, \bibinfo {author} {\bibfnamefont
  {A.}~\bibnamefont {{Bulgarelli}}}, \bibinfo {author} {\bibfnamefont
  {M.}~\bibnamefont {{Capaccioli}}}, \bibinfo {author} {\bibfnamefont
  {P.}~\bibnamefont {{Casella}}}, \bibinfo {author} {\bibfnamefont {A.~J.}\
  \bibnamefont {{Castro-Tirado}}}, \bibinfo {author} {\bibfnamefont
  {E.}~\bibnamefont {{Chassande-Mottin}}}, \bibinfo {author} {\bibfnamefont
  {R.}~\bibnamefont {{Ciolfi}}}, \bibinfo {author} {\bibfnamefont {C.~M.}\
  \bibnamefont {{Copperwheat}}}, \bibinfo {author} {\bibfnamefont
  {M.}~\bibnamefont {{Dadina}}}, \bibinfo {author} {\bibfnamefont
  {G.}~\bibnamefont {{De Cesare}}}, \bibinfo {author} {\bibfnamefont
  {A.}~\bibnamefont {{di Paola}}}, \bibinfo {author} {\bibfnamefont {Y.~Z.}\
  \bibnamefont {{Fan}}}, \bibinfo {author} {\bibfnamefont {B.}~\bibnamefont
  {{Gendre}}}, \bibinfo {author} {\bibfnamefont {G.}~\bibnamefont
  {{Giuffrida}}}, \bibinfo {author} {\bibfnamefont {A.}~\bibnamefont
  {{Giunta}}}, \bibinfo {author} {\bibfnamefont {L.~K.}\ \bibnamefont
  {{Hunt}}}, \bibinfo {author} {\bibfnamefont {G.~L.}\ \bibnamefont
  {{Israel}}}, \bibinfo {author} {\bibfnamefont {Z.-P.}\ \bibnamefont {{Jin}}},
  \bibinfo {author} {\bibfnamefont {M.~M.}\ \bibnamefont {{Kasliwal}}},
  \bibinfo {author} {\bibfnamefont {S.}~\bibnamefont {{Klose}}}, \bibinfo
  {author} {\bibfnamefont {M.}~\bibnamefont {{Lisi}}}, \bibinfo {author}
  {\bibfnamefont {F.}~\bibnamefont {{Longo}}}, \bibinfo {author} {\bibfnamefont
  {E.}~\bibnamefont {{Maiorano}}}, \bibinfo {author} {\bibfnamefont
  {M.}~\bibnamefont {{Mapelli}}}, \bibinfo {author} {\bibfnamefont
  {N.}~\bibnamefont {{Masetti}}}, \bibinfo {author} {\bibfnamefont
  {L.}~\bibnamefont {{Nava}}}, \bibinfo {author} {\bibfnamefont
  {B.}~\bibnamefont {{Patricelli}}}, \bibinfo {author} {\bibfnamefont
  {D.}~\bibnamefont {{Perley}}}, \bibinfo {author} {\bibfnamefont
  {A.}~\bibnamefont {{Pescalli}}}, \bibinfo {author} {\bibfnamefont
  {T.}~\bibnamefont {{Piran}}}, \bibinfo {author} {\bibfnamefont
  {A.}~\bibnamefont {{Possenti}}}, \bibinfo {author} {\bibfnamefont
  {L.}~\bibnamefont {{Pulone}}}, \bibinfo {author} {\bibfnamefont
  {M.}~\bibnamefont {{Razzano}}}, \bibinfo {author} {\bibfnamefont
  {R.}~\bibnamefont {{Salvaterra}}}, \bibinfo {author} {\bibfnamefont
  {P.}~\bibnamefont {{Schipani}}}, \bibinfo {author} {\bibfnamefont
  {M.}~\bibnamefont {{Spera}}}, \bibinfo {author} {\bibfnamefont
  {A.}~\bibnamefont {{Stamerra}}}, \bibinfo {author} {\bibfnamefont
  {L.}~\bibnamefont {{Stella}}}, \bibinfo {author} {\bibfnamefont
  {G.}~\bibnamefont {{Tagliaferri}}}, \bibinfo {author} {\bibfnamefont
  {V.}~\bibnamefont {{Testa}}}, \bibinfo {author} {\bibfnamefont
  {E.}~\bibnamefont {{Troja}}}, \bibinfo {author} {\bibfnamefont
  {M.}~\bibnamefont {{Turatto}}}, \bibinfo {author} {\bibfnamefont {S.~D.}\
  \bibnamefont {{Vergani}}},\ and\ \bibinfo {author} {\bibfnamefont
  {D.}~\bibnamefont {{Vergani}}},\ }\bibfield  {title} {\bibinfo {title}
  {{Spectroscopic identification of r-process nucleosynthesis in a double
  neutron-star merger}},\ }\href {https://doi.org/10.1038/nature24298}
  {\bibfield  {journal} {\bibinfo  {journal} {"Nature"}\ }\textbf {\bibinfo
  {volume} {551}},\ \bibinfo {pages} {67} (\bibinfo {year} {2017})},\ \Eprint
  {https://arxiv.org/abs/1710.05858} {arXiv:1710.05858 [astro-ph.HE]}
  \BibitemShut {NoStop}%
\bibitem [{\citenamefont {{Lattimer}}\ and\ \citenamefont
  {{Schramm}}(1976)}]{1976ApJ...210..549L}%
  \BibitemOpen
  \bibfield  {author} {\bibinfo {author} {\bibfnamefont {J.~M.}\ \bibnamefont
  {{Lattimer}}}\ and\ \bibinfo {author} {\bibfnamefont {D.~N.}\ \bibnamefont
  {{Schramm}}},\ }\bibfield  {title} {\bibinfo {title} {{The tidal disruption
  of neutron stars by black holes in close binaries.}},\ }\href
  {https://doi.org/10.1086/154860} {\bibfield  {journal} {\bibinfo  {journal}
  {Astroph.J.}\ }\textbf {\bibinfo {volume} {210}},\ \bibinfo {pages} {549}
  (\bibinfo {year} {1976})}\BibitemShut {NoStop}%
\bibitem [{\citenamefont {{Roberts}}\ \emph {et~al.}(2011)\citenamefont
  {{Roberts}}, \citenamefont {{Kasen}}, \citenamefont {{Lee}},\ and\
  \citenamefont {{Ramirez-Ruiz}}}]{2011ApJ...736L..21R}%
  \BibitemOpen
  \bibfield  {author} {\bibinfo {author} {\bibfnamefont {L.~F.}\ \bibnamefont
  {{Roberts}}}, \bibinfo {author} {\bibfnamefont {D.}~\bibnamefont {{Kasen}}},
  \bibinfo {author} {\bibfnamefont {W.~H.}\ \bibnamefont {{Lee}}},\ and\
  \bibinfo {author} {\bibfnamefont {E.}~\bibnamefont {{Ramirez-Ruiz}}},\
  }\bibfield  {title} {\bibinfo {title} {{Electromagnetic Transients Powered by
  Nuclear Decay in the Tidal Tails of Coalescing Compact Binaries}},\ }\href
  {https://doi.org/10.1088/2041-8205/736/1/L21} {\bibfield  {journal} {\bibinfo
   {journal} {Astroph.J.Letters}\ }\textbf {\bibinfo {volume} {736}},\ \bibinfo
  {eid} {L21} (\bibinfo {year} {2011})},\ \Eprint
  {https://arxiv.org/abs/1104.5504} {arXiv:1104.5504 [astro-ph.HE]}
  \BibitemShut {NoStop}%
\bibitem [{\citenamefont {Baiotti}\ and\ \citenamefont
  {Rezzolla}(2017)}]{Baiotti:2016qnr}%
  \BibitemOpen
  \bibfield  {author} {\bibinfo {author} {\bibfnamefont {L.}~\bibnamefont
  {Baiotti}}\ and\ \bibinfo {author} {\bibfnamefont {L.}~\bibnamefont
  {Rezzolla}},\ }\bibfield  {title} {\bibinfo {title} {{Binary neutron star
  mergers: a review of Einstein\textquoteright{}s richest laboratory}},\ }\href
  {https://doi.org/10.1088/1361-6633/aa67bb} {\bibfield  {journal} {\bibinfo
  {journal} {Rept. Prog. Phys.}\ }\textbf {\bibinfo {volume} {80}},\ \bibinfo
  {pages} {096901} (\bibinfo {year} {2017})},\ \Eprint
  {https://arxiv.org/abs/1607.03540} {arXiv:1607.03540 [gr-qc]} \BibitemShut
  {NoStop}%
\bibitem [{\citenamefont {Burns}(2020)}]{Burns:2019byj}%
  \BibitemOpen
  \bibfield  {author} {\bibinfo {author} {\bibfnamefont {E.}~\bibnamefont
  {Burns}},\ }\bibfield  {title} {\bibinfo {title} {{Neutron Star Mergers and
  How to Study Them}},\ }\href {https://doi.org/10.1007/s41114-020-00028-7}
  {\bibfield  {journal} {\bibinfo  {journal} {Living Rev. Rel.}\ }\textbf
  {\bibinfo {volume} {23}},\ \bibinfo {pages} {4} (\bibinfo {year} {2020})},\
  \Eprint {https://arxiv.org/abs/1909.06085} {arXiv:1909.06085 [astro-ph.HE]}
  \BibitemShut {NoStop}%
\bibitem [{\citenamefont {Foucart}(2020)}]{Foucart:2020ats}%
  \BibitemOpen
  \bibfield  {author} {\bibinfo {author} {\bibfnamefont {F.}~\bibnamefont
  {Foucart}},\ }\bibfield  {title} {\bibinfo {title} {{A brief overview of
  black hole-neutron star mergers}},\ }\href
  {https://doi.org/10.3389/fspas.2020.00046} {\bibfield  {journal} {\bibinfo
  {journal} {Front. Astron. Space Sci.}\ }\textbf {\bibinfo {volume} {7}},\
  \bibinfo {pages} {46} (\bibinfo {year} {2020})},\ \Eprint
  {https://arxiv.org/abs/2006.10570} {arXiv:2006.10570 [astro-ph.HE]}
  \BibitemShut {NoStop}%
\bibitem [{\citenamefont {Radice}\ \emph {et~al.}(2020)\citenamefont {Radice},
  \citenamefont {Bernuzzi},\ and\ \citenamefont {Perego}}]{Radice:2020ddv}%
  \BibitemOpen
  \bibfield  {author} {\bibinfo {author} {\bibfnamefont {D.}~\bibnamefont
  {Radice}}, \bibinfo {author} {\bibfnamefont {S.}~\bibnamefont {Bernuzzi}},\
  and\ \bibinfo {author} {\bibfnamefont {A.}~\bibnamefont {Perego}},\
  }\bibfield  {title} {\bibinfo {title} {{The Dynamics of Binary Neutron Star
  Mergers and GW170817}},\ }\href
  {https://doi.org/10.1146/annurev-nucl-013120-114541} {\bibfield  {journal}
  {\bibinfo  {journal} {Ann. Rev. Nucl. Part. Sci.}\ }\textbf {\bibinfo
  {volume} {70}},\ \bibinfo {pages} {95} (\bibinfo {year} {2020})},\ \Eprint
  {https://arxiv.org/abs/2002.03863} {arXiv:2002.03863 [astro-ph.HE]}
  \BibitemShut {NoStop}%
\bibitem [{\citenamefont {Kyutoku}\ \emph {et~al.}(2021)\citenamefont
  {Kyutoku}, \citenamefont {Shibata},\ and\ \citenamefont
  {Taniguchi}}]{Kyutoku:2021icp}%
  \BibitemOpen
  \bibfield  {author} {\bibinfo {author} {\bibfnamefont {K.}~\bibnamefont
  {Kyutoku}}, \bibinfo {author} {\bibfnamefont {M.}~\bibnamefont {Shibata}},\
  and\ \bibinfo {author} {\bibfnamefont {K.}~\bibnamefont {Taniguchi}},\
  }\bibfield  {title} {\bibinfo {title} {{Coalescence of black
  hole\textendash{}neutron star binaries}},\ }\href
  {https://doi.org/10.1007/s41114-021-00033-4} {\bibfield  {journal} {\bibinfo
  {journal} {Living Rev. Rel.}\ }\textbf {\bibinfo {volume} {24}},\ \bibinfo
  {pages} {5} (\bibinfo {year} {2021})},\ \Eprint
  {https://arxiv.org/abs/2110.06218} {arXiv:2110.06218 [astro-ph.HE]}
  \BibitemShut {NoStop}%
\bibitem [{\citenamefont {Barnes}\ and\ \citenamefont
  {Kasen}(2013)}]{Barnes:2013wka}%
  \BibitemOpen
  \bibfield  {author} {\bibinfo {author} {\bibfnamefont {J.}~\bibnamefont
  {Barnes}}\ and\ \bibinfo {author} {\bibfnamefont {D.}~\bibnamefont {Kasen}},\
  }\bibfield  {title} {\bibinfo {title} {{Effect of a High Opacity on the Light
  Curves of Radioactively Powered Transients from Compact Object Mergers}},\
  }\href {https://doi.org/10.1088/0004-637X/775/1/18} {\bibfield  {journal}
  {\bibinfo  {journal} {Astrophys. J.}\ }\textbf {\bibinfo {volume} {775}},\
  \bibinfo {pages} {18} (\bibinfo {year} {2013})},\ \Eprint
  {https://arxiv.org/abs/1303.5787} {arXiv:1303.5787 [astro-ph.HE]}
  \BibitemShut {NoStop}%
\bibitem [{\citenamefont {Alford}\ \emph {et~al.}(2008)\citenamefont {Alford},
  \citenamefont {Schmitt}, \citenamefont {Rajagopal},\ and\ \citenamefont
  {Schafer}}]{RevModPhys.80.1455}%
  \BibitemOpen
  \bibfield  {author} {\bibinfo {author} {\bibfnamefont {M.~G.}\ \bibnamefont
  {Alford}}, \bibinfo {author} {\bibfnamefont {A.}~\bibnamefont {Schmitt}},
  \bibinfo {author} {\bibfnamefont {K.}~\bibnamefont {Rajagopal}},\ and\
  \bibinfo {author} {\bibfnamefont {T.}~\bibnamefont {Schafer}},\ }\bibfield
  {title} {\bibinfo {title} {Color superconductivity in dense quark matter},\
  }\href {https://doi.org/10.1103/RevModPhys.80.1455} {\bibfield  {journal}
  {\bibinfo  {journal} {Rev. Mod. Phys.}\ }\textbf {\bibinfo {volume} {80}},\
  \bibinfo {pages} {1455} (\bibinfo {year} {2008})}\BibitemShut {NoStop}%
\bibitem [{\citenamefont {Buballa}\ \emph {et~al.}(2020)\citenamefont
  {Buballa}, \citenamefont {Carignano},\ and\ \citenamefont
  {Kurth}}]{Buballa:2020xaa}%
  \BibitemOpen
  \bibfield  {author} {\bibinfo {author} {\bibfnamefont {M.}~\bibnamefont
  {Buballa}}, \bibinfo {author} {\bibfnamefont {S.}~\bibnamefont {Carignano}},\
  and\ \bibinfo {author} {\bibfnamefont {L.}~\bibnamefont {Kurth}},\ }\bibfield
   {title} {\bibinfo {title} {{Inhomogeneous phases in the quark-meson model
  with explicit chiral-symmetry breaking}},\ }\href
  {https://doi.org/10.1140/epjst/e2020-000101-x} {\bibfield  {journal}
  {\bibinfo  {journal} {Eur. Phys. J. ST}\ }\textbf {\bibinfo {volume} {229}},\
  \bibinfo {pages} {3371} (\bibinfo {year} {2020})},\ \Eprint
  {https://arxiv.org/abs/2006.02133} {arXiv:2006.02133 [hep-ph]} \BibitemShut
  {NoStop}%
\bibitem [{\citenamefont {Abbott}\ \emph {et~al.}(2018)\citenamefont {Abbott}
  \emph {et~al.}}]{GW170817-NSRadius}%
  \BibitemOpen
  \bibfield  {author} {\bibinfo {author} {\bibfnamefont {B.}~\bibnamefont
  {Abbott}} \emph {et~al.} (\bibinfo {collaboration} {LIGO Scientific,
  Virgo}),\ }\bibfield  {title} {\bibinfo {title} {{GW170817: Measurements of
  neutron star radii and equation of state}},\ }\href
  {https://doi.org/10.1103/PhysRevLett.121.161101} {\bibfield  {journal}
  {\bibinfo  {journal} {Phys. Rev. Lett.}\ }\textbf {\bibinfo {volume} {121}},\
  \bibinfo {pages} {161101} (\bibinfo {year} {2018})},\ \Eprint
  {https://arxiv.org/abs/1805.11581} {arXiv:1805.11581 [gr-qc]} \BibitemShut
  {NoStop}%
\bibitem [{\citenamefont {Radice}\ \emph {et~al.}(2018)\citenamefont {Radice},
  \citenamefont {Perego}, \citenamefont {Zappa},\ and\ \citenamefont
  {Bernuzzi}}]{2017ApJL2041}%
  \BibitemOpen
  \bibfield  {author} {\bibinfo {author} {\bibfnamefont {D.}~\bibnamefont
  {Radice}}, \bibinfo {author} {\bibfnamefont {A.}~\bibnamefont {Perego}},
  \bibinfo {author} {\bibfnamefont {F.}~\bibnamefont {Zappa}},\ and\ \bibinfo
  {author} {\bibfnamefont {S.}~\bibnamefont {Bernuzzi}},\ }\bibfield  {title}
  {\bibinfo {title} {Gw170817: Joint constraint on the neutron star equation of
  state from multimessenger observations},\ }\href
  {http://stacks.iop.org/2041-8205/852/i=2/a=L29} {\bibfield  {journal}
  {\bibinfo  {journal} {The Astrophysical Journal Letters}\ }\textbf {\bibinfo
  {volume} {852}},\ \bibinfo {pages} {L29} (\bibinfo {year} {2018})},\ \Eprint
  {https://arxiv.org/abs/1711.03647} {arXiv:1711.03647 [astro-ph.HE]}
  \BibitemShut {NoStop}%
\bibitem [{\citenamefont {Raaijmakers}\ \emph {et~al.}(2021)\citenamefont
  {Raaijmakers}, \citenamefont {Greif}, \citenamefont {Hebeler}, \citenamefont
  {Hinderer}, \citenamefont {Nissanke}, \citenamefont {Schwenk}, \citenamefont
  {Riley}, \citenamefont {Watts}, \citenamefont {Lattimer},\ and\ \citenamefont
  {Ho}}]{Raaijmakers:2021uju}%
  \BibitemOpen
  \bibfield  {author} {\bibinfo {author} {\bibfnamefont {G.}~\bibnamefont
  {Raaijmakers}}, \bibinfo {author} {\bibfnamefont {S.~K.}\ \bibnamefont
  {Greif}}, \bibinfo {author} {\bibfnamefont {K.}~\bibnamefont {Hebeler}},
  \bibinfo {author} {\bibfnamefont {T.}~\bibnamefont {Hinderer}}, \bibinfo
  {author} {\bibfnamefont {S.}~\bibnamefont {Nissanke}}, \bibinfo {author}
  {\bibfnamefont {A.}~\bibnamefont {Schwenk}}, \bibinfo {author} {\bibfnamefont
  {T.~E.}\ \bibnamefont {Riley}}, \bibinfo {author} {\bibfnamefont {A.~L.}\
  \bibnamefont {Watts}}, \bibinfo {author} {\bibfnamefont {J.~M.}\ \bibnamefont
  {Lattimer}},\ and\ \bibinfo {author} {\bibfnamefont {W.~C.~G.}\ \bibnamefont
  {Ho}},\ }\bibfield  {title} {\bibinfo {title} {{Constraints on the Dense
  Matter Equation of State and Neutron Star Properties from
  NICER\textquoteright{}s Mass\textendash{}Radius Estimate of PSR J0740+6620
  and Multimessenger Observations}},\ }\href
  {https://doi.org/10.3847/2041-8213/ac089a} {\bibfield  {journal} {\bibinfo
  {journal} {Astrophys. J. Lett.}\ }\textbf {\bibinfo {volume} {918}},\
  \bibinfo {pages} {L29} (\bibinfo {year} {2021})},\ \Eprint
  {https://arxiv.org/abs/2105.06981} {arXiv:2105.06981 [astro-ph.HE]}
  \BibitemShut {NoStop}%
\bibitem [{\citenamefont {Miller}\ \emph {et~al.}(2021)\citenamefont {Miller}
  \emph {et~al.}}]{Miller:2021qha}%
  \BibitemOpen
  \bibfield  {author} {\bibinfo {author} {\bibfnamefont {M.~C.}\ \bibnamefont
  {Miller}} \emph {et~al.},\ }\bibfield  {title} {\bibinfo {title} {{The Radius
  of PSR J0740+6620 from NICER and XMM-Newton Data}},\ }\href
  {https://doi.org/10.3847/2041-8213/ac089b} {\bibfield  {journal} {\bibinfo
  {journal} {Astrophys. J. Lett.}\ }\textbf {\bibinfo {volume} {918}},\
  \bibinfo {pages} {L28} (\bibinfo {year} {2021})},\ \Eprint
  {https://arxiv.org/abs/2105.06979} {arXiv:2105.06979 [astro-ph.HE]}
  \BibitemShut {NoStop}%
\bibitem [{\citenamefont {Horowitz}\ \emph {et~al.}(2014)\citenamefont
  {Horowitz}, \citenamefont {Kumar},\ and\ \citenamefont
  {Michaels}}]{Horowitz:2013wha}%
  \BibitemOpen
  \bibfield  {author} {\bibinfo {author} {\bibfnamefont {C.~J.}\ \bibnamefont
  {Horowitz}}, \bibinfo {author} {\bibfnamefont {K.~S.}\ \bibnamefont
  {Kumar}},\ and\ \bibinfo {author} {\bibfnamefont {R.}~\bibnamefont
  {Michaels}},\ }\bibfield  {title} {\bibinfo {title} {{Electroweak
  Measurements of Neutron Densities in CREX and PREX at JLab, USA}},\ }\href
  {https://doi.org/10.1140/epja/i2014-14048-3} {\bibfield  {journal} {\bibinfo
  {journal} {Eur. Phys. J.}\ }\textbf {\bibinfo {volume} {A50}},\ \bibinfo
  {pages} {48} (\bibinfo {year} {2014})},\ \Eprint
  {https://arxiv.org/abs/1307.3572} {arXiv:1307.3572 [nucl-ex]} \BibitemShut
  {NoStop}%
\bibitem [{\citenamefont {Reed}\ \emph {et~al.}(2021)\citenamefont {Reed},
  \citenamefont {Fattoyev}, \citenamefont {Horowitz},\ and\ \citenamefont
  {Piekarewicz}}]{PhysRevLett.126.172503}%
  \BibitemOpen
  \bibfield  {author} {\bibinfo {author} {\bibfnamefont {B.~T.}\ \bibnamefont
  {Reed}}, \bibinfo {author} {\bibfnamefont {F.~J.}\ \bibnamefont {Fattoyev}},
  \bibinfo {author} {\bibfnamefont {C.~J.}\ \bibnamefont {Horowitz}},\ and\
  \bibinfo {author} {\bibfnamefont {J.}~\bibnamefont {Piekarewicz}},\
  }\bibfield  {title} {\bibinfo {title} {Implications of prex-2 on the equation
  of state of neutron-rich matter},\ }\href
  {https://doi.org/10.1103/PhysRevLett.126.172503} {\bibfield  {journal}
  {\bibinfo  {journal} {Phys. Rev. Lett.}\ }\textbf {\bibinfo {volume} {126}},\
  \bibinfo {pages} {172503} (\bibinfo {year} {2021})}\BibitemShut {NoStop}%
\bibitem [{\citenamefont {Kajino}\ \emph {et~al.}(2019)\citenamefont {Kajino},
  \citenamefont {Aoki}, \citenamefont {Balantekin}, \citenamefont {Diehl},
  \citenamefont {Famiano},\ and\ \citenamefont {Mathews}}]{Kajino:2019abv}%
  \BibitemOpen
  \bibfield  {author} {\bibinfo {author} {\bibfnamefont {T.}~\bibnamefont
  {Kajino}}, \bibinfo {author} {\bibfnamefont {W.}~\bibnamefont {Aoki}},
  \bibinfo {author} {\bibfnamefont {A.~B.}\ \bibnamefont {Balantekin}},
  \bibinfo {author} {\bibfnamefont {R.}~\bibnamefont {Diehl}}, \bibinfo
  {author} {\bibfnamefont {M.~A.}\ \bibnamefont {Famiano}},\ and\ \bibinfo
  {author} {\bibfnamefont {G.~J.}\ \bibnamefont {Mathews}},\ }\bibfield
  {title} {\bibinfo {title} {{Current status of r -process nucleosynthesis}},\
  }\href {https://doi.org/10.1016/j.ppnp.2019.02.008} {\bibfield  {journal}
  {\bibinfo  {journal} {Prog. Part. Nucl. Phys.}\ }\textbf {\bibinfo {volume}
  {107}},\ \bibinfo {pages} {109} (\bibinfo {year} {2019})},\ \Eprint
  {https://arxiv.org/abs/1906.05002} {arXiv:1906.05002 [astro-ph.HE]}
  \BibitemShut {NoStop}%
\bibitem [{\citenamefont {Siegel}(2022)}]{Siegel:2022upa}%
  \BibitemOpen
  \bibfield  {author} {\bibinfo {author} {\bibfnamefont {D.~M.}\ \bibnamefont
  {Siegel}},\ }\bibfield  {title} {\bibinfo {title} {{r-Process nucleosynthesis
  in gravitational-wave and other explosive astrophysical events}},\ }\href
  {https://doi.org/10.1038/s42254-022-00439-1} {\bibfield  {journal} {\bibinfo
  {journal} {Nature Rev. Phys.}\ }\textbf {\bibinfo {volume} {4}},\ \bibinfo
  {pages} {306} (\bibinfo {year} {2022})}\BibitemShut {NoStop}%
\bibitem [{\citenamefont {Fern\'andez}\ \emph {et~al.}(2020)\citenamefont
  {Fern\'andez}, \citenamefont {Foucart},\ and\ \citenamefont
  {Lippuner}}]{Fernandez:2020oow}%
  \BibitemOpen
  \bibfield  {author} {\bibinfo {author} {\bibfnamefont {R.}~\bibnamefont
  {Fern\'andez}}, \bibinfo {author} {\bibfnamefont {F.}~\bibnamefont
  {Foucart}},\ and\ \bibinfo {author} {\bibfnamefont {J.}~\bibnamefont
  {Lippuner}},\ }\bibfield  {title} {\bibinfo {title} {{The landscape of disc
  outflows from black hole\textendash{}neutron star mergers}},\ }\href
  {https://doi.org/10.1093/mnras/staa2209} {\bibfield  {journal} {\bibinfo
  {journal} {Mon. Not. Roy. Astron. Soc.}\ }\textbf {\bibinfo {volume} {497}},\
  \bibinfo {pages} {3221} (\bibinfo {year} {2020})},\ \Eprint
  {https://arxiv.org/abs/2005.14208} {arXiv:2005.14208 [astro-ph.HE]}
  \BibitemShut {NoStop}%
\bibitem [{\citenamefont {Surman}\ \emph {et~al.}(2006)\citenamefont {Surman},
  \citenamefont {McLaughlin},\ and\ \citenamefont {Hix}}]{Surman:2005kf}%
  \BibitemOpen
  \bibfield  {author} {\bibinfo {author} {\bibfnamefont {R.}~\bibnamefont
  {Surman}}, \bibinfo {author} {\bibfnamefont {G.~C.}\ \bibnamefont
  {McLaughlin}},\ and\ \bibinfo {author} {\bibfnamefont {W.~R.}\ \bibnamefont
  {Hix}},\ }\bibfield  {title} {\bibinfo {title} {{Nucleosynthesis in the
  outflow from gamma-ray burst accretion disks}},\ }\href
  {https://doi.org/10.1086/501116} {\bibfield  {journal} {\bibinfo  {journal}
  {Astrophys. J.}\ }\textbf {\bibinfo {volume} {643}},\ \bibinfo {pages} {1057}
  (\bibinfo {year} {2006})},\ \Eprint {https://arxiv.org/abs/astro-ph/0509365}
  {arXiv:astro-ph/0509365} \BibitemShut {NoStop}%
\bibitem [{\citenamefont {{Lippuner}}\ and\ \citenamefont
  {{Roberts}}(2015)}]{Lippuner2015}%
  \BibitemOpen
  \bibfield  {author} {\bibinfo {author} {\bibfnamefont {J.}~\bibnamefont
  {{Lippuner}}}\ and\ \bibinfo {author} {\bibfnamefont {L.~F.}\ \bibnamefont
  {{Roberts}}},\ }\bibfield  {title} {\bibinfo {title} {{r-process Lanthanide
  Production and Heating Rates in Kilonovae}},\ }\href
  {https://doi.org/10.1088/0004-637X/815/2/82} {\bibfield  {journal} {\bibinfo
  {journal} {Astroph.J.}\ }\textbf {\bibinfo {volume} {815}},\ \bibinfo {eid}
  {82} (\bibinfo {year} {2015})},\ \Eprint {https://arxiv.org/abs/1508.03133}
  {arXiv:1508.03133 [astro-ph.HE]} \BibitemShut {NoStop}%
\bibitem [{\citenamefont {Holmbeck}\ \emph {et~al.}(2020)\citenamefont
  {Holmbeck}, \citenamefont {Surman}, \citenamefont {Frebel}, \citenamefont
  {McLaughlin}, \citenamefont {Mumpower}, \citenamefont {Sprouse},
  \citenamefont {Kawano}, \citenamefont {Vassh},\ and\ \citenamefont
  {Beers}}]{Holmbeck:2020qlp}%
  \BibitemOpen
  \bibfield  {author} {\bibinfo {author} {\bibfnamefont {E.~M.}\ \bibnamefont
  {Holmbeck}}, \bibinfo {author} {\bibfnamefont {R.}~\bibnamefont {Surman}},
  \bibinfo {author} {\bibfnamefont {A.}~\bibnamefont {Frebel}}, \bibinfo
  {author} {\bibfnamefont {G.~C.}\ \bibnamefont {McLaughlin}}, \bibinfo
  {author} {\bibfnamefont {M.~R.}\ \bibnamefont {Mumpower}}, \bibinfo {author}
  {\bibfnamefont {T.~M.}\ \bibnamefont {Sprouse}}, \bibinfo {author}
  {\bibfnamefont {T.}~\bibnamefont {Kawano}}, \bibinfo {author} {\bibfnamefont
  {N.}~\bibnamefont {Vassh}},\ and\ \bibinfo {author} {\bibfnamefont {T.~C.}\
  \bibnamefont {Beers}},\ }\bibfield  {title} {\bibinfo {title}
  {{Characterizing $r$-Process Sites through Actinide Production}},\ }\href
  {https://doi.org/10.1088/1742-6596/1668/1/012020} {\bibfield  {journal}
  {\bibinfo  {journal} {J. Phys. Conf. Ser.}\ }\textbf {\bibinfo {volume}
  {1668}},\ \bibinfo {pages} {012020} (\bibinfo {year} {2020})},\ \Eprint
  {https://arxiv.org/abs/2001.08792} {arXiv:2001.08792 [astro-ph.HE]}
  \BibitemShut {NoStop}%
\bibitem [{\citenamefont {Vogel}(1984)}]{PhysRevD.29.1918}%
  \BibitemOpen
  \bibfield  {author} {\bibinfo {author} {\bibfnamefont {P.}~\bibnamefont
  {Vogel}},\ }\bibfield  {title} {\bibinfo {title} {Analysis of the
  antineutrino capture on protons},\ }\href
  {https://doi.org/10.1103/PhysRevD.29.1918} {\bibfield  {journal} {\bibinfo
  {journal} {Phys. Rev. D}\ }\textbf {\bibinfo {volume} {29}},\ \bibinfo
  {pages} {1918} (\bibinfo {year} {1984})}\BibitemShut {NoStop}%
\bibitem [{\citenamefont {{Bruenn}}(1985)}]{1985ApJS...58..771B}%
  \BibitemOpen
  \bibfield  {author} {\bibinfo {author} {\bibfnamefont {S.~W.}\ \bibnamefont
  {{Bruenn}}},\ }\bibfield  {title} {\bibinfo {title} {{Stellar core collapse -
  Numerical model and infall epoch}},\ }\href {https://doi.org/10.1086/191056}
  {\bibfield  {journal} {\bibinfo  {journal} {Astroph.J.Suppl.}\ }\textbf
  {\bibinfo {volume} {58}},\ \bibinfo {pages} {771} (\bibinfo {year}
  {1985})}\BibitemShut {NoStop}%
\bibitem [{\citenamefont {{Salmonson}}\ and\ \citenamefont
  {{Wilson}}(1999)}]{1999ApJ...517..859S}%
  \BibitemOpen
  \bibfield  {author} {\bibinfo {author} {\bibfnamefont {J.~D.}\ \bibnamefont
  {{Salmonson}}}\ and\ \bibinfo {author} {\bibfnamefont {J.~R.}\ \bibnamefont
  {{Wilson}}},\ }\bibfield  {title} {\bibinfo {title} {{General Relativistic
  Augmentation of Neutrino Pair Annihilation Energy Deposition nea\ r Neutron
  Stars}},\ }\href {https://doi.org/10.1086/307232} {\bibfield  {journal}
  {\bibinfo  {journal} {Astroph.J.}\ }\textbf {\bibinfo {volume} {517}},\
  \bibinfo {pages} {859} (\bibinfo {year} {1999})},\ \Eprint
  {https://arxiv.org/abs/astro-ph/9908017} {astro-ph/9908017} \BibitemShut
  {NoStop}%
\bibitem [{\citenamefont {Burrows}\ \emph {et~al.}(2006)\citenamefont
  {Burrows}, \citenamefont {Reddy},\ and\ \citenamefont
  {Thompson}}]{Burrows:2004vq}%
  \BibitemOpen
  \bibfield  {author} {\bibinfo {author} {\bibfnamefont {A.}~\bibnamefont
  {Burrows}}, \bibinfo {author} {\bibfnamefont {S.}~\bibnamefont {Reddy}},\
  and\ \bibinfo {author} {\bibfnamefont {T.~A.}\ \bibnamefont {Thompson}},\
  }\bibfield  {title} {\bibinfo {title} {{Neutrino opacities in nuclear
  matter}},\ }\href {https://doi.org/10.1016/j.nuclphysa.2004.06.012}
  {\bibfield  {journal} {\bibinfo  {journal} {Nucl. Phys. A}\ }\textbf
  {\bibinfo {volume} {777}},\ \bibinfo {pages} {356} (\bibinfo {year}
  {2006})},\ \Eprint {https://arxiv.org/abs/astro-ph/0404432}
  {arXiv:astro-ph/0404432} \BibitemShut {NoStop}%
\bibitem [{\citenamefont {Foucart}(2023)}]{Foucart:2022bth}%
  \BibitemOpen
  \bibfield  {author} {\bibinfo {author} {\bibfnamefont {F.}~\bibnamefont
  {Foucart}},\ }\bibfield  {title} {\bibinfo {title} {{Neutrino transport in
  general relativistic neutron star merger simulations}},\ }\href
  {https://doi.org/10.1007/s41115-023-00016-y} {\bibfield  {journal} {\bibinfo
  {journal} {Liv. Rev. Comput. Astrophys.}\ }\textbf {\bibinfo {volume} {9}},\
  \bibinfo {pages} {1} (\bibinfo {year} {2023})},\ \Eprint
  {https://arxiv.org/abs/2209.02538} {arXiv:2209.02538 [astro-ph.HE]}
  \BibitemShut {NoStop}%
\bibitem [{\citenamefont {Loffredo}\ \emph {et~al.}(2023)\citenamefont
  {Loffredo}, \citenamefont {Perego}, \citenamefont {Logoteta},\ and\
  \citenamefont {Branchesi}}]{Loffredo:2022prq}%
  \BibitemOpen
  \bibfield  {author} {\bibinfo {author} {\bibfnamefont {E.}~\bibnamefont
  {Loffredo}}, \bibinfo {author} {\bibfnamefont {A.}~\bibnamefont {Perego}},
  \bibinfo {author} {\bibfnamefont {D.}~\bibnamefont {Logoteta}},\ and\
  \bibinfo {author} {\bibfnamefont {M.}~\bibnamefont {Branchesi}},\ }\bibfield
  {title} {\bibinfo {title} {{Muons in the aftermath of neutron star mergers
  and their impact on trapped neutrinos}},\ }\href
  {https://doi.org/10.1051/0004-6361/202244927} {\bibfield  {journal} {\bibinfo
   {journal} {Astron. Astrophys.}\ }\textbf {\bibinfo {volume} {672}},\
  \bibinfo {pages} {A124} (\bibinfo {year} {2023})},\ \Eprint
  {https://arxiv.org/abs/2209.04458} {arXiv:2209.04458 [astro-ph.HE]}
  \BibitemShut {NoStop}%
\bibitem [{\citenamefont {Foucart}\ \emph {et~al.}(2016)\citenamefont
  {Foucart}, \citenamefont {O'Connor}, \citenamefont {Roberts}, \citenamefont
  {Kidder}, \citenamefont {Pfeiffer},\ and\ \citenamefont
  {Scheel}}]{Foucart:2016rxm}%
  \BibitemOpen
  \bibfield  {author} {\bibinfo {author} {\bibfnamefont {F.}~\bibnamefont
  {Foucart}}, \bibinfo {author} {\bibfnamefont {E.}~\bibnamefont {O'Connor}},
  \bibinfo {author} {\bibfnamefont {L.}~\bibnamefont {Roberts}}, \bibinfo
  {author} {\bibfnamefont {L.~E.}\ \bibnamefont {Kidder}}, \bibinfo {author}
  {\bibfnamefont {H.~P.}\ \bibnamefont {Pfeiffer}},\ and\ \bibinfo {author}
  {\bibfnamefont {M.~A.}\ \bibnamefont {Scheel}},\ }\bibfield  {title}
  {\bibinfo {title} {{Impact of an improved neutrino energy estimate on
  outflows in neutron star merger simulations}},\ }\href
  {https://doi.org/10.1103/PhysRevD.94.123016} {\bibfield  {journal} {\bibinfo
  {journal} {Phys. Rev. D}\ }\textbf {\bibinfo {volume} {94}},\ \bibinfo
  {pages} {123016} (\bibinfo {year} {2016})},\ \Eprint
  {https://arxiv.org/abs/1607.07450} {arXiv:1607.07450 [astro-ph.HE]}
  \BibitemShut {NoStop}%
\bibitem [{\citenamefont {Bahcall}\ and\ \citenamefont
  {Davis}(1976)}]{Bahcall:1976zz}%
  \BibitemOpen
  \bibfield  {author} {\bibinfo {author} {\bibfnamefont {J.~N.}\ \bibnamefont
  {Bahcall}}\ and\ \bibinfo {author} {\bibfnamefont {R.}~\bibnamefont
  {Davis}},\ }\bibfield  {title} {\bibinfo {title} {{Solar Neutrinos - a
  Scientific Puzzle}},\ }\href {https://doi.org/10.1126/science.191.4224.264}
  {\bibfield  {journal} {\bibinfo  {journal} {Science}\ }\textbf {\bibinfo
  {volume} {191}},\ \bibinfo {pages} {264} (\bibinfo {year}
  {1976})}\BibitemShut {NoStop}%
\bibitem [{\citenamefont {Wolfenstein}(1978)}]{Wolfenstein:1977ue}%
  \BibitemOpen
  \bibfield  {author} {\bibinfo {author} {\bibfnamefont {L.}~\bibnamefont
  {Wolfenstein}},\ }\bibfield  {title} {\bibinfo {title} {{Neutrino
  Oscillations in Matter}},\ }\href {https://doi.org/10.1103/PhysRevD.17.2369}
  {\bibfield  {journal} {\bibinfo  {journal} {Phys. Rev. D}\ }\textbf {\bibinfo
  {volume} {17}},\ \bibinfo {pages} {2369} (\bibinfo {year}
  {1978})}\BibitemShut {NoStop}%
\bibitem [{\citenamefont {Bilenky}\ and\ \citenamefont
  {Pontecorvo}(1978)}]{Bilenky:1978nj}%
  \BibitemOpen
  \bibfield  {author} {\bibinfo {author} {\bibfnamefont {S.~M.}\ \bibnamefont
  {Bilenky}}\ and\ \bibinfo {author} {\bibfnamefont {B.}~\bibnamefont
  {Pontecorvo}},\ }\bibfield  {title} {\bibinfo {title} {{Lepton Mixing and
  Neutrino Oscillations}},\ }\href
  {https://doi.org/10.1016/0370-1573(78)90095-9} {\bibfield  {journal}
  {\bibinfo  {journal} {Phys. Rept.}\ }\textbf {\bibinfo {volume} {41}},\
  \bibinfo {pages} {225} (\bibinfo {year} {1978})}\BibitemShut {NoStop}%
\bibitem [{\citenamefont {Mikheyev}\ and\ \citenamefont
  {Smirnov}(1985)}]{Mikheyev:1985zog}%
  \BibitemOpen
  \bibfield  {author} {\bibinfo {author} {\bibfnamefont {S.~P.}\ \bibnamefont
  {Mikheyev}}\ and\ \bibinfo {author} {\bibfnamefont {A.~Y.}\ \bibnamefont
  {Smirnov}},\ }\bibfield  {title} {\bibinfo {title} {{Resonance Amplification
  of Oscillations in Matter and Spectroscopy of Solar Neutrinos}},\ }\href@noop
  {} {\bibfield  {journal} {\bibinfo  {journal} {Sov. J. Nucl. Phys.}\ }\textbf
  {\bibinfo {volume} {42}},\ \bibinfo {pages} {913} (\bibinfo {year}
  {1985})}\BibitemShut {NoStop}%
\bibitem [{\citenamefont {Kyutoku}\ and\ \citenamefont
  {Kashiyama}(2018)}]{Kyutoku:2017wnb}%
  \BibitemOpen
  \bibfield  {author} {\bibinfo {author} {\bibfnamefont {K.}~\bibnamefont
  {Kyutoku}}\ and\ \bibinfo {author} {\bibfnamefont {K.}~\bibnamefont
  {Kashiyama}},\ }\bibfield  {title} {\bibinfo {title} {{Detectability of
  thermal neutrinos from binary-neutron-star mergers and implication to
  neutrino physics}},\ }\href {https://doi.org/10.1103/PhysRevD.97.103001}
  {\bibfield  {journal} {\bibinfo  {journal} {Phys. Rev. D}\ }\textbf {\bibinfo
  {volume} {97}},\ \bibinfo {pages} {103001} (\bibinfo {year} {2018})},\
  \Eprint {https://arxiv.org/abs/1710.05922} {arXiv:1710.05922 [astro-ph.HE]}
  \BibitemShut {NoStop}%
\bibitem [{\citenamefont {{Caballero}}\ \emph {et~al.}(2014)\citenamefont
  {{Caballero}}, \citenamefont {{Malkus}}, \citenamefont {{McLaughlin}},\ and\
  \citenamefont {{Surman}}}]{2014JPhG...41d4004C}%
  \BibitemOpen
  \bibfield  {author} {\bibinfo {author} {\bibfnamefont {O.~L.}\ \bibnamefont
  {{Caballero}}}, \bibinfo {author} {\bibfnamefont {A.~C.}\ \bibnamefont
  {{Malkus}}}, \bibinfo {author} {\bibfnamefont {G.~C.}\ \bibnamefont
  {{McLaughlin}}},\ and\ \bibinfo {author} {\bibfnamefont {R.~A.}\ \bibnamefont
  {{Surman}}},\ }\bibfield  {title} {\bibinfo {title} {{The influence of
  neutrinos on the nucleosynthesis of accretion disc outflows}},\ }\href
  {https://doi.org/10.1088/0954-3899/41/4/044004} {\bibfield  {journal}
  {\bibinfo  {journal} {Journal of Physics G Nuclear Physics}\ }\textbf
  {\bibinfo {volume} {41}},\ \bibinfo {eid} {044004} (\bibinfo {year}
  {2014})}\BibitemShut {NoStop}%
\bibitem [{\citenamefont {Zhu}\ \emph {et~al.}(2016)\citenamefont {Zhu},
  \citenamefont {Perego},\ and\ \citenamefont {McLaughlin}}]{Zhu:2016mwa}%
  \BibitemOpen
  \bibfield  {author} {\bibinfo {author} {\bibfnamefont {Y.-L.}\ \bibnamefont
  {Zhu}}, \bibinfo {author} {\bibfnamefont {A.}~\bibnamefont {Perego}},\ and\
  \bibinfo {author} {\bibfnamefont {G.~C.}\ \bibnamefont {McLaughlin}},\
  }\bibfield  {title} {\bibinfo {title} {{Matter Neutrino Resonance Transitions
  above a Neutron Star Merger Remnant}},\ }\href
  {https://doi.org/10.1103/PhysRevD.94.105006} {\bibfield  {journal} {\bibinfo
  {journal} {Phys. Rev. D}\ }\textbf {\bibinfo {volume} {94}},\ \bibinfo
  {pages} {105006} (\bibinfo {year} {2016})},\ \Eprint
  {https://arxiv.org/abs/1607.04671} {arXiv:1607.04671 [hep-ph]} \BibitemShut
  {NoStop}%
\bibitem [{\citenamefont {Padilla-Gay}\ \emph {et~al.}(2024)\citenamefont
  {Padilla-Gay}, \citenamefont {Shalgar},\ and\ \citenamefont
  {Tamborra}}]{Padilla-Gay:2024wyo}%
  \BibitemOpen
  \bibfield  {author} {\bibinfo {author} {\bibfnamefont {I.}~\bibnamefont
  {Padilla-Gay}}, \bibinfo {author} {\bibfnamefont {S.}~\bibnamefont
  {Shalgar}},\ and\ \bibinfo {author} {\bibfnamefont {I.}~\bibnamefont
  {Tamborra}},\ }\bibfield  {title} {\bibinfo {title} {{Symmetry breaking due
  to multi-angle matter-neutrino resonance in neutron star merger remnants}},\
  }\href {https://doi.org/10.1088/1475-7516/2024/05/037} {\bibfield  {journal}
  {\bibinfo  {journal} {JCAP}\ }\textbf {\bibinfo {volume} {05}},\ \bibinfo
  {pages} {037}},\ \Eprint {https://arxiv.org/abs/2403.15532} {arXiv:2403.15532
  [astro-ph.HE]} \BibitemShut {NoStop}%
\bibitem [{\citenamefont {Banerjee}\ \emph {et~al.}(2011)\citenamefont
  {Banerjee}, \citenamefont {Dighe},\ and\ \citenamefont
  {Raffelt}}]{PhysRevD.84.053013}%
  \BibitemOpen
  \bibfield  {author} {\bibinfo {author} {\bibfnamefont {A.}~\bibnamefont
  {Banerjee}}, \bibinfo {author} {\bibfnamefont {A.}~\bibnamefont {Dighe}},\
  and\ \bibinfo {author} {\bibfnamefont {G.}~\bibnamefont {Raffelt}},\
  }\bibfield  {title} {\bibinfo {title} {Linearized flavor-stability analysis
  of dense neutrino streams},\ }\href
  {https://doi.org/10.1103/PhysRevD.84.053013} {\bibfield  {journal} {\bibinfo
  {journal} {Phys. Rev. D}\ }\textbf {\bibinfo {volume} {84}},\ \bibinfo
  {pages} {053013} (\bibinfo {year} {2011})}\BibitemShut {NoStop}%
\bibitem [{\citenamefont {Wu}\ \emph {et~al.}(2017)\citenamefont {Wu},
  \citenamefont {Tamborra}, \citenamefont {Just},\ and\ \citenamefont
  {Janka}}]{Wu:2017drk}%
  \BibitemOpen
  \bibfield  {author} {\bibinfo {author} {\bibfnamefont {M.-R.}\ \bibnamefont
  {Wu}}, \bibinfo {author} {\bibfnamefont {I.}~\bibnamefont {Tamborra}},
  \bibinfo {author} {\bibfnamefont {O.}~\bibnamefont {Just}},\ and\ \bibinfo
  {author} {\bibfnamefont {H.-T.}\ \bibnamefont {Janka}},\ }\bibfield  {title}
  {\bibinfo {title} {{Imprints of neutrino-pair flavor conversions on
  nucleosynthesis in ejecta from neutron-star merger remnants}},\ }\href
  {https://doi.org/10.1103/PhysRevD.96.123015} {\bibfield  {journal} {\bibinfo
  {journal} {Phys. Rev. D}\ }\textbf {\bibinfo {volume} {96}},\ \bibinfo
  {pages} {123015} (\bibinfo {year} {2017})},\ \Eprint
  {https://arxiv.org/abs/1711.00477} {arXiv:1711.00477 [astro-ph.HE]}
  \BibitemShut {NoStop}%
\bibitem [{\citenamefont {Wu}\ and\ \citenamefont
  {Tamborra}(2017)}]{Wu:2017qpc}%
  \BibitemOpen
  \bibfield  {author} {\bibinfo {author} {\bibfnamefont {M.-R.}\ \bibnamefont
  {Wu}}\ and\ \bibinfo {author} {\bibfnamefont {I.}~\bibnamefont {Tamborra}},\
  }\bibfield  {title} {\bibinfo {title} {{Fast neutrino conversions: Ubiquitous
  in compact binary merger remnants}},\ }\href
  {https://doi.org/10.1103/PhysRevD.95.103007} {\bibfield  {journal} {\bibinfo
  {journal} {Phys. Rev. D}\ }\textbf {\bibinfo {volume} {95}},\ \bibinfo
  {pages} {103007} (\bibinfo {year} {2017})},\ \Eprint
  {https://arxiv.org/abs/1701.06580} {arXiv:1701.06580 [astro-ph.HE]}
  \BibitemShut {NoStop}%
\bibitem [{\citenamefont {Grohs}\ \emph {et~al.}(2023)\citenamefont {Grohs},
  \citenamefont {Richers}, \citenamefont {Couch}, \citenamefont {Foucart},
  \citenamefont {Kneller},\ and\ \citenamefont {McLaughlin}}]{Grohs:2022fyq}%
  \BibitemOpen
  \bibfield  {author} {\bibinfo {author} {\bibfnamefont {E.}~\bibnamefont
  {Grohs}}, \bibinfo {author} {\bibfnamefont {S.}~\bibnamefont {Richers}},
  \bibinfo {author} {\bibfnamefont {S.~M.}\ \bibnamefont {Couch}}, \bibinfo
  {author} {\bibfnamefont {F.}~\bibnamefont {Foucart}}, \bibinfo {author}
  {\bibfnamefont {J.~P.}\ \bibnamefont {Kneller}},\ and\ \bibinfo {author}
  {\bibfnamefont {G.~C.}\ \bibnamefont {McLaughlin}},\ }\bibfield  {title}
  {\bibinfo {title} {{Neutrino fast flavor instability in three dimensions for
  a neutron star merger}},\ }\href
  {https://doi.org/10.1016/j.physletb.2023.138210} {\bibfield  {journal}
  {\bibinfo  {journal} {Phys. Lett. B}\ }\textbf {\bibinfo {volume} {846}},\
  \bibinfo {pages} {138210} (\bibinfo {year} {2023})},\ \Eprint
  {https://arxiv.org/abs/2207.02214} {arXiv:2207.02214 [hep-ph]} \BibitemShut
  {NoStop}%
\bibitem [{\citenamefont {Izaguirre}\ \emph {et~al.}(2017)\citenamefont
  {Izaguirre}, \citenamefont {Raffelt},\ and\ \citenamefont
  {Tamborra}}]{Izaguirre:2016gsx}%
  \BibitemOpen
  \bibfield  {author} {\bibinfo {author} {\bibfnamefont {I.}~\bibnamefont
  {Izaguirre}}, \bibinfo {author} {\bibfnamefont {G.}~\bibnamefont {Raffelt}},\
  and\ \bibinfo {author} {\bibfnamefont {I.}~\bibnamefont {Tamborra}},\
  }\bibfield  {title} {\bibinfo {title} {{Fast Pairwise Conversion of Supernova
  Neutrinos: A Dispersion-Relation Approach}},\ }\href
  {https://doi.org/10.1103/PhysRevLett.118.021101} {\bibfield  {journal}
  {\bibinfo  {journal} {Phys. Rev. Lett.}\ }\textbf {\bibinfo {volume} {118}},\
  \bibinfo {pages} {021101} (\bibinfo {year} {2017})},\ \Eprint
  {https://arxiv.org/abs/1610.01612} {arXiv:1610.01612 [hep-ph]} \BibitemShut
  {NoStop}%
\bibitem [{\citenamefont {Morinaga}(2022)}]{Morinaga:2021vmc}%
  \BibitemOpen
  \bibfield  {author} {\bibinfo {author} {\bibfnamefont {T.}~\bibnamefont
  {Morinaga}},\ }\bibfield  {title} {\bibinfo {title} {{Fast neutrino flavor
  instability and neutrino flavor lepton number crossings}},\ }\href
  {https://doi.org/10.1103/PhysRevD.105.L101301} {\bibfield  {journal}
  {\bibinfo  {journal} {Phys. Rev. D}\ }\textbf {\bibinfo {volume} {105}},\
  \bibinfo {pages} {L101301} (\bibinfo {year} {2022})},\ \Eprint
  {https://arxiv.org/abs/2103.15267} {arXiv:2103.15267 [hep-ph]} \BibitemShut
  {NoStop}%
\bibitem [{\citenamefont {Johns}\ and\ \citenamefont
  {Nagakura}(2021)}]{Johns:2021taz}%
  \BibitemOpen
  \bibfield  {author} {\bibinfo {author} {\bibfnamefont {L.}~\bibnamefont
  {Johns}}\ and\ \bibinfo {author} {\bibfnamefont {H.}~\bibnamefont
  {Nagakura}},\ }\bibfield  {title} {\bibinfo {title} {{Fast flavor
  instabilities and the search for neutrino angular crossings}},\ }\href
  {https://doi.org/10.1103/PhysRevD.103.123012} {\bibfield  {journal} {\bibinfo
   {journal} {Phys. Rev. D}\ }\textbf {\bibinfo {volume} {103}},\ \bibinfo
  {pages} {123012} (\bibinfo {year} {2021})},\ \Eprint
  {https://arxiv.org/abs/2104.04106} {arXiv:2104.04106 [hep-ph]} \BibitemShut
  {NoStop}%
\bibitem [{\citenamefont {Fiorillo}\ and\ \citenamefont
  {Raffelt}(2024{\natexlab{a}})}]{Fiorillo:2024bzm}%
  \BibitemOpen
  \bibfield  {author} {\bibinfo {author} {\bibfnamefont {D.~F.~G.}\
  \bibnamefont {Fiorillo}}\ and\ \bibinfo {author} {\bibfnamefont {G.~G.}\
  \bibnamefont {Raffelt}},\ }\bibfield  {title} {\bibinfo {title} {{Theory of
  neutrino fast flavor evolution. Part I. Linear response theory and stability
  conditions.}},\ }\href {https://doi.org/10.1007/JHEP08(2024)225} {\bibfield
  {journal} {\bibinfo  {journal} {J. High Energy Phys.}\ }\textbf {\bibinfo
  {volume} {08}},\ \bibinfo {pages} {225}},\ \Eprint
  {https://arxiv.org/abs/2406.06708} {arXiv:2406.06708 [hep-ph]} \BibitemShut
  {NoStop}%
\bibitem [{\citenamefont {Fiorillo}\ and\ \citenamefont
  {Raffelt}(2024{\natexlab{b}})}]{Fiorillo:2024qbl}%
  \BibitemOpen
  \bibfield  {author} {\bibinfo {author} {\bibfnamefont {D.~F.~G.}\
  \bibnamefont {Fiorillo}}\ and\ \bibinfo {author} {\bibfnamefont
  {G.}~\bibnamefont {Raffelt}},\ }\bibfield  {title} {\bibinfo {title} {{Fast
  flavor conversions at the edge of instability}},\ }\href@noop {} {\bibfield
  {journal} {\bibinfo  {journal} {arXiv}\ } (\bibinfo {year}
  {2024}{\natexlab{b}})},\ \Eprint {https://arxiv.org/abs/2403.12189}
  {arXiv:2403.12189 [hep-ph]} \BibitemShut {NoStop}%
\bibitem [{\citenamefont {Li}\ and\ \citenamefont {Siegel}(2021)}]{Li:2021vqj}%
  \BibitemOpen
  \bibfield  {author} {\bibinfo {author} {\bibfnamefont {X.}~\bibnamefont
  {Li}}\ and\ \bibinfo {author} {\bibfnamefont {D.~M.}\ \bibnamefont
  {Siegel}},\ }\bibfield  {title} {\bibinfo {title} {{Neutrino Fast Flavor
  Conversions in Neutron-Star Postmerger Accretion Disks}},\ }\href
  {https://doi.org/10.1103/PhysRevLett.126.251101} {\bibfield  {journal}
  {\bibinfo  {journal} {Phys. Rev. Lett.}\ }\textbf {\bibinfo {volume} {126}},\
  \bibinfo {pages} {251101} (\bibinfo {year} {2021})},\ \Eprint
  {https://arxiv.org/abs/2103.02616} {arXiv:2103.02616 [astro-ph.HE]}
  \BibitemShut {NoStop}%
\bibitem [{\citenamefont {Fern\'andez}\ \emph {et~al.}(2022)\citenamefont
  {Fern\'andez}, \citenamefont {Richers}, \citenamefont {Mulyk},\ and\
  \citenamefont {Fahlman}}]{Fernandez:2022yyv}%
  \BibitemOpen
  \bibfield  {author} {\bibinfo {author} {\bibfnamefont {R.}~\bibnamefont
  {Fern\'andez}}, \bibinfo {author} {\bibfnamefont {S.}~\bibnamefont
  {Richers}}, \bibinfo {author} {\bibfnamefont {N.}~\bibnamefont {Mulyk}},\
  and\ \bibinfo {author} {\bibfnamefont {S.}~\bibnamefont {Fahlman}},\
  }\bibfield  {title} {\bibinfo {title} {{Fast flavor instability in
  hypermassive neutron star disk outflows}},\ }\href
  {https://doi.org/10.1103/PhysRevD.106.103003} {\bibfield  {journal} {\bibinfo
   {journal} {Phys. Rev. D}\ }\textbf {\bibinfo {volume} {106}},\ \bibinfo
  {pages} {103003} (\bibinfo {year} {2022})},\ \Eprint
  {https://arxiv.org/abs/2207.10680} {arXiv:2207.10680 [astro-ph.HE]}
  \BibitemShut {NoStop}%
\bibitem [{\citenamefont {Just}\ \emph {et~al.}(2022)\citenamefont {Just},
  \citenamefont {Abbar}, \citenamefont {Wu}, \citenamefont {Tamborra},
  \citenamefont {Janka},\ and\ \citenamefont {Capozzi}}]{Just:2022flt}%
  \BibitemOpen
  \bibfield  {author} {\bibinfo {author} {\bibfnamefont {O.}~\bibnamefont
  {Just}}, \bibinfo {author} {\bibfnamefont {S.}~\bibnamefont {Abbar}},
  \bibinfo {author} {\bibfnamefont {M.-R.}\ \bibnamefont {Wu}}, \bibinfo
  {author} {\bibfnamefont {I.}~\bibnamefont {Tamborra}}, \bibinfo {author}
  {\bibfnamefont {H.-T.}\ \bibnamefont {Janka}},\ and\ \bibinfo {author}
  {\bibfnamefont {F.}~\bibnamefont {Capozzi}},\ }\bibfield  {title} {\bibinfo
  {title} {{Fast neutrino conversion in hydrodynamic simulations of
  neutrino-cooled accretion disks}},\ }\href
  {https://doi.org/10.1103/PhysRevD.105.083024} {\bibfield  {journal} {\bibinfo
   {journal} {Phys. Rev. D}\ }\textbf {\bibinfo {volume} {105}},\ \bibinfo
  {pages} {083024} (\bibinfo {year} {2022})},\ \Eprint
  {https://arxiv.org/abs/2203.16559} {arXiv:2203.16559 [astro-ph.HE]}
  \BibitemShut {NoStop}%
\bibitem [{\citenamefont {Johns}\ and\ \citenamefont
  {Xiong}(2022)}]{Johns:2022yqy}%
  \BibitemOpen
  \bibfield  {author} {\bibinfo {author} {\bibfnamefont {L.}~\bibnamefont
  {Johns}}\ and\ \bibinfo {author} {\bibfnamefont {Z.}~\bibnamefont {Xiong}},\
  }\bibfield  {title} {\bibinfo {title} {{Collisional instabilities of
  neutrinos and their interplay with fast flavor conversion in compact
  objects}},\ }\href {https://doi.org/10.1103/PhysRevD.106.103029} {\bibfield
  {journal} {\bibinfo  {journal} {Phys. Rev. D}\ }\textbf {\bibinfo {volume}
  {106}},\ \bibinfo {pages} {103029} (\bibinfo {year} {2022})},\ \Eprint
  {https://arxiv.org/abs/2208.11059} {arXiv:2208.11059 [hep-ph]} \BibitemShut
  {NoStop}%
\bibitem [{\citenamefont {Akaho}\ \emph {et~al.}(2024)\citenamefont {Akaho},
  \citenamefont {Liu}, \citenamefont {Nagakura}, \citenamefont {Zaizen},\ and\
  \citenamefont {Yamada}}]{Akaho:2023brj}%
  \BibitemOpen
  \bibfield  {author} {\bibinfo {author} {\bibfnamefont {R.}~\bibnamefont
  {Akaho}}, \bibinfo {author} {\bibfnamefont {J.}~\bibnamefont {Liu}}, \bibinfo
  {author} {\bibfnamefont {H.}~\bibnamefont {Nagakura}}, \bibinfo {author}
  {\bibfnamefont {M.}~\bibnamefont {Zaizen}},\ and\ \bibinfo {author}
  {\bibfnamefont {S.}~\bibnamefont {Yamada}},\ }\bibfield  {title} {\bibinfo
  {title} {{Collisional and fast neutrino flavor instabilities in
  two-dimensional core-collapse supernova simulation with Boltzmann neutrino
  transport}},\ }\href {https://doi.org/10.1103/PhysRevD.109.023012} {\bibfield
   {journal} {\bibinfo  {journal} {Phys. Rev. D}\ }\textbf {\bibinfo {volume}
  {109}},\ \bibinfo {pages} {023012} (\bibinfo {year} {2024})},\ \Eprint
  {https://arxiv.org/abs/2311.11272} {arXiv:2311.11272 [astro-ph.HE]}
  \BibitemShut {NoStop}%
\bibitem [{\citenamefont {Shalgar}\ and\ \citenamefont
  {Tamborra}(2024)}]{Shalgar:2024gjt}%
  \BibitemOpen
  \bibfield  {author} {\bibinfo {author} {\bibfnamefont {S.}~\bibnamefont
  {Shalgar}}\ and\ \bibinfo {author} {\bibfnamefont {I.}~\bibnamefont
  {Tamborra}},\ }\bibfield  {title} {\bibinfo {title} {{Neutrino quantum
  kinetics in a core-collapse supernova}},\ }\href
  {https://doi.org/10.1088/1475-7516/2024/09/021} {\bibfield  {journal}
  {\bibinfo  {journal} {JCAP}\ }\textbf {\bibinfo {volume} {09}},\ \bibinfo
  {pages} {021}},\ \Eprint {https://arxiv.org/abs/2406.09504} {arXiv:2406.09504
  [astro-ph.HE]} \BibitemShut {NoStop}%
\bibitem [{\citenamefont {Tamborra}\ and\ \citenamefont
  {Shalgar}(2021)}]{Tamborra:2020cul}%
  \BibitemOpen
  \bibfield  {author} {\bibinfo {author} {\bibfnamefont {I.}~\bibnamefont
  {Tamborra}}\ and\ \bibinfo {author} {\bibfnamefont {S.}~\bibnamefont
  {Shalgar}},\ }\bibfield  {title} {\bibinfo {title} {{New Developments in
  Flavor Evolution of a Dense Neutrino Gas}},\ }\href
  {https://doi.org/10.1146/annurev-nucl-102920-050505} {\bibfield  {journal}
  {\bibinfo  {journal} {Ann. Rev. Nucl. Part. Sci.}\ }\textbf {\bibinfo
  {volume} {71}},\ \bibinfo {pages} {165} (\bibinfo {year} {2021})},\ \Eprint
  {https://arxiv.org/abs/2011.01948} {arXiv:2011.01948 [astro-ph.HE]}
  \BibitemShut {NoStop}%
\bibitem [{\citenamefont {Richers}\ and\ \citenamefont
  {Sen}(2022)}]{Richers:2022zug}%
  \BibitemOpen
  \bibfield  {author} {\bibinfo {author} {\bibfnamefont {S.}~\bibnamefont
  {Richers}}\ and\ \bibinfo {author} {\bibfnamefont {M.}~\bibnamefont {Sen}},\
  }\bibinfo {title} {{Fast Flavor Transformations}},\ in\ \href@noop {} {\emph
  {\bibinfo {booktitle} {{Handbook of Nuclear Physics}}}},\ \bibinfo {editor}
  {edited by\ \bibinfo {editor} {\bibfnamefont {I.}~\bibnamefont {Tanihata}},
  \bibinfo {editor} {\bibfnamefont {H.}~\bibnamefont {Toki}},\ and\ \bibinfo
  {editor} {\bibfnamefont {T.}~\bibnamefont {Kajino}}}\ (\bibinfo {year}
  {2022})\ pp.\ \bibinfo {pages} {1--17},\ \Eprint
  {https://arxiv.org/abs/2207.03561} {arXiv:2207.03561 [astro-ph.HE]}
  \BibitemShut {NoStop}%
\bibitem [{\citenamefont {Fujibayashi}\ \emph {et~al.}(2020)\citenamefont
  {Fujibayashi}, \citenamefont {Wanajo}, \citenamefont {Kiuchi}, \citenamefont
  {Kyutoku}, \citenamefont {Sekiguchi},\ and\ \citenamefont
  {Shibata}}]{Fujibayashi:2020dvr}%
  \BibitemOpen
  \bibfield  {author} {\bibinfo {author} {\bibfnamefont {S.}~\bibnamefont
  {Fujibayashi}}, \bibinfo {author} {\bibfnamefont {S.}~\bibnamefont {Wanajo}},
  \bibinfo {author} {\bibfnamefont {K.}~\bibnamefont {Kiuchi}}, \bibinfo
  {author} {\bibfnamefont {K.}~\bibnamefont {Kyutoku}}, \bibinfo {author}
  {\bibfnamefont {Y.}~\bibnamefont {Sekiguchi}},\ and\ \bibinfo {author}
  {\bibfnamefont {M.}~\bibnamefont {Shibata}},\ }\bibfield  {title} {\bibinfo
  {title} {{Postmerger Mass Ejection of Low-mass Binary Neutron Stars}},\
  }\href {https://doi.org/10.3847/1538-4357/abafc2} {\bibfield  {journal}
  {\bibinfo  {journal} {Astrophys. J.}\ }\textbf {\bibinfo {volume} {901}},\
  \bibinfo {pages} {122} (\bibinfo {year} {2020})},\ \Eprint
  {https://arxiv.org/abs/2007.00474} {arXiv:2007.00474 [astro-ph.HE]}
  \BibitemShut {NoStop}%
\bibitem [{\citenamefont {Deaton}\ \emph {et~al.}(2013)\citenamefont {Deaton},
  \citenamefont {Duez}, \citenamefont {Foucart}, \citenamefont {O'Connor},
  \citenamefont {Ott}, \citenamefont {Kidder}, \citenamefont {Muhlberger},
  \citenamefont {Scheel},\ and\ \citenamefont {Szilagyi}}]{Deaton:2013sla}%
  \BibitemOpen
  \bibfield  {author} {\bibinfo {author} {\bibfnamefont {M.~B.}\ \bibnamefont
  {Deaton}}, \bibinfo {author} {\bibfnamefont {M.~D.}\ \bibnamefont {Duez}},
  \bibinfo {author} {\bibfnamefont {F.}~\bibnamefont {Foucart}}, \bibinfo
  {author} {\bibfnamefont {E.}~\bibnamefont {O'Connor}}, \bibinfo {author}
  {\bibfnamefont {C.~D.}\ \bibnamefont {Ott}}, \bibinfo {author} {\bibfnamefont
  {L.~E.}\ \bibnamefont {Kidder}}, \bibinfo {author} {\bibfnamefont {C.~D.}\
  \bibnamefont {Muhlberger}}, \bibinfo {author} {\bibfnamefont {M.~A.}\
  \bibnamefont {Scheel}},\ and\ \bibinfo {author} {\bibfnamefont
  {B.}~\bibnamefont {Szilagyi}},\ }\bibfield  {title} {\bibinfo {title} {{Black
  Hole-Neutron Star Mergers with a Hot Nuclear Equation of State: Outflow and
  Neutrino-Cooled Disk for a Low-Mass, High-Spin Case}},\ }\href
  {https://doi.org/10.1088/0004-637X/776/1/47} {\bibfield  {journal} {\bibinfo
  {journal} {Astrophys. J.}\ }\textbf {\bibinfo {volume} {776}},\ \bibinfo
  {pages} {47} (\bibinfo {year} {2013})},\ \Eprint
  {https://arxiv.org/abs/1304.3384} {arXiv:1304.3384 [astro-ph.HE]}
  \BibitemShut {NoStop}%
\bibitem [{\citenamefont {Sekiguchi}\ \emph {et~al.}(2016)\citenamefont
  {Sekiguchi}, \citenamefont {Kiuchi}, \citenamefont {Kyutoku}, \citenamefont
  {Shibata},\ and\ \citenamefont {Taniguchi}}]{Sekiguchi:2016bjd}%
  \BibitemOpen
  \bibfield  {author} {\bibinfo {author} {\bibfnamefont {Y.}~\bibnamefont
  {Sekiguchi}}, \bibinfo {author} {\bibfnamefont {K.}~\bibnamefont {Kiuchi}},
  \bibinfo {author} {\bibfnamefont {K.}~\bibnamefont {Kyutoku}}, \bibinfo
  {author} {\bibfnamefont {M.}~\bibnamefont {Shibata}},\ and\ \bibinfo {author}
  {\bibfnamefont {K.}~\bibnamefont {Taniguchi}},\ }\bibfield  {title} {\bibinfo
  {title} {{Dynamical mass ejection from the merger of asymmetric binary
  neutron stars: Radiation-hydrodynamics study in general relativity}},\ }\href
  {https://doi.org/10.1103/PhysRevD.93.124046} {\bibfield  {journal} {\bibinfo
  {journal} {Phys. Rev. D}\ }\textbf {\bibinfo {volume} {93}},\ \bibinfo
  {pages} {124046} (\bibinfo {year} {2016})},\ \Eprint
  {https://arxiv.org/abs/1603.01918} {arXiv:1603.01918 [astro-ph.HE]}
  \BibitemShut {NoStop}%
\bibitem [{\citenamefont {Radice}\ \emph {et~al.}(2022)\citenamefont {Radice},
  \citenamefont {Bernuzzi}, \citenamefont {Perego},\ and\ \citenamefont
  {Haas}}]{Radice:2021jtw}%
  \BibitemOpen
  \bibfield  {author} {\bibinfo {author} {\bibfnamefont {D.}~\bibnamefont
  {Radice}}, \bibinfo {author} {\bibfnamefont {S.}~\bibnamefont {Bernuzzi}},
  \bibinfo {author} {\bibfnamefont {A.}~\bibnamefont {Perego}},\ and\ \bibinfo
  {author} {\bibfnamefont {R.}~\bibnamefont {Haas}},\ }\bibfield  {title}
  {\bibinfo {title} {{A new moment-based general-relativistic
  neutrino-radiation transport code: Methods and first applications to neutron
  star mergers}},\ }\href {https://doi.org/10.1093/mnras/stac589} {\bibfield
  {journal} {\bibinfo  {journal} {Mon. Not. Roy. Astron. Soc.}\ }\textbf
  {\bibinfo {volume} {512}},\ \bibinfo {pages} {1499} (\bibinfo {year}
  {2022})},\ \Eprint {https://arxiv.org/abs/2111.14858} {arXiv:2111.14858
  [astro-ph.HE]} \BibitemShut {NoStop}%
\bibitem [{\citenamefont {Shibata}\ \emph {et~al.}(2021)\citenamefont
  {Shibata}, \citenamefont {Fujibayashi},\ and\ \citenamefont
  {Sekiguchi}}]{Shibata:2021xmo}%
  \BibitemOpen
  \bibfield  {author} {\bibinfo {author} {\bibfnamefont {M.}~\bibnamefont
  {Shibata}}, \bibinfo {author} {\bibfnamefont {S.}~\bibnamefont
  {Fujibayashi}},\ and\ \bibinfo {author} {\bibfnamefont {Y.}~\bibnamefont
  {Sekiguchi}},\ }\bibfield  {title} {\bibinfo {title} {{Long-term evolution of
  neutron-star merger remnants in general relativistic resistive
  magnetohydrodynamics with a mean-field dynamo term}},\ }\href
  {https://doi.org/10.1103/PhysRevD.104.063026} {\bibfield  {journal} {\bibinfo
   {journal} {Phys. Rev. D}\ }\textbf {\bibinfo {volume} {104}},\ \bibinfo
  {pages} {063026} (\bibinfo {year} {2021})},\ \Eprint
  {https://arxiv.org/abs/2109.08732} {arXiv:2109.08732 [astro-ph.HE]}
  \BibitemShut {NoStop}%
\bibitem [{\citenamefont {Fujibayashi}\ \emph {et~al.}(2023)\citenamefont
  {Fujibayashi}, \citenamefont {Kiuchi}, \citenamefont {Wanajo}, \citenamefont
  {Kyutoku}, \citenamefont {Sekiguchi},\ and\ \citenamefont
  {Shibata}}]{Fujibayashi:2022ftg}%
  \BibitemOpen
  \bibfield  {author} {\bibinfo {author} {\bibfnamefont {S.}~\bibnamefont
  {Fujibayashi}}, \bibinfo {author} {\bibfnamefont {K.}~\bibnamefont {Kiuchi}},
  \bibinfo {author} {\bibfnamefont {S.}~\bibnamefont {Wanajo}}, \bibinfo
  {author} {\bibfnamefont {K.}~\bibnamefont {Kyutoku}}, \bibinfo {author}
  {\bibfnamefont {Y.}~\bibnamefont {Sekiguchi}},\ and\ \bibinfo {author}
  {\bibfnamefont {M.}~\bibnamefont {Shibata}},\ }\bibfield  {title} {\bibinfo
  {title} {{Comprehensive Study of Mass Ejection and Nucleosynthesis in Binary
  Neutron Star Mergers Leaving Short-lived Massive Neutron Stars}},\ }\href
  {https://doi.org/10.3847/1538-4357/ac9ce0} {\bibfield  {journal} {\bibinfo
  {journal} {Astrophys. J.}\ }\textbf {\bibinfo {volume} {942}},\ \bibinfo
  {pages} {39} (\bibinfo {year} {2023})},\ \Eprint
  {https://arxiv.org/abs/2205.05557} {arXiv:2205.05557 [astro-ph.HE]}
  \BibitemShut {NoStop}%
\bibitem [{\citenamefont {Fern\'andez}\ \emph {et~al.}(2019)\citenamefont
  {Fern\'andez}, \citenamefont {Tchekhovskoy}, \citenamefont {Quataert},
  \citenamefont {Foucart},\ and\ \citenamefont {Kasen}}]{Fernandez:2018kax}%
  \BibitemOpen
  \bibfield  {author} {\bibinfo {author} {\bibfnamefont {R.}~\bibnamefont
  {Fern\'andez}}, \bibinfo {author} {\bibfnamefont {A.}~\bibnamefont
  {Tchekhovskoy}}, \bibinfo {author} {\bibfnamefont {E.}~\bibnamefont
  {Quataert}}, \bibinfo {author} {\bibfnamefont {F.}~\bibnamefont {Foucart}},\
  and\ \bibinfo {author} {\bibfnamefont {D.}~\bibnamefont {Kasen}},\ }\bibfield
   {title} {\bibinfo {title} {{Long-term GRMHD simulations of neutron star
  merger accretion discs: implications for electromagnetic counterparts}},\
  }\href {https://doi.org/10.1093/mnras/sty2932} {\bibfield  {journal}
  {\bibinfo  {journal} {Mon. Not. Roy. Astron. Soc.}\ }\textbf {\bibinfo
  {volume} {482}},\ \bibinfo {pages} {3373} (\bibinfo {year} {2019})},\ \Eprint
  {https://arxiv.org/abs/1808.00461} {arXiv:1808.00461 [astro-ph.HE]}
  \BibitemShut {NoStop}%
\bibitem [{\citenamefont {Just}\ \emph {et~al.}(2021)\citenamefont {Just},
  \citenamefont {Goriely}, \citenamefont {Janka}, \citenamefont {Nagataki},\
  and\ \citenamefont {Bauswein}}]{Just:2021cls}%
  \BibitemOpen
  \bibfield  {author} {\bibinfo {author} {\bibfnamefont {O.}~\bibnamefont
  {Just}}, \bibinfo {author} {\bibfnamefont {S.}~\bibnamefont {Goriely}},
  \bibinfo {author} {\bibfnamefont {H.-T.}\ \bibnamefont {Janka}}, \bibinfo
  {author} {\bibfnamefont {S.}~\bibnamefont {Nagataki}},\ and\ \bibinfo
  {author} {\bibfnamefont {A.}~\bibnamefont {Bauswein}},\ }\bibfield  {title}
  {\bibinfo {title} {{Neutrino absorption and other physics dependencies in
  neutrino-cooled black hole accretion discs}},\ }\href
  {https://doi.org/10.1093/mnras/stab2861} {\bibfield  {journal} {\bibinfo
  {journal} {Mon. Not. Roy. Astron. Soc.}\ }\textbf {\bibinfo {volume} {509}},\
  \bibinfo {pages} {1377} (\bibinfo {year} {2021})},\ \Eprint
  {https://arxiv.org/abs/2102.08387} {arXiv:2102.08387 [astro-ph.HE]}
  \BibitemShut {NoStop}%
\bibitem [{\citenamefont {De}\ and\ \citenamefont {Siegel}(2021)}]{De:2020jdt}%
  \BibitemOpen
  \bibfield  {author} {\bibinfo {author} {\bibfnamefont {S.}~\bibnamefont
  {De}}\ and\ \bibinfo {author} {\bibfnamefont {D.~M.}\ \bibnamefont
  {Siegel}},\ }\bibfield  {title} {\bibinfo {title} {{Igniting Weak
  Interactions in Neutron Star Postmerger Accretion Disks}},\ }\href
  {https://doi.org/10.3847/1538-4357/ac110b} {\bibfield  {journal} {\bibinfo
  {journal} {Astrophys. J.}\ }\textbf {\bibinfo {volume} {921}},\ \bibinfo
  {pages} {94} (\bibinfo {year} {2021})},\ \Eprint
  {https://arxiv.org/abs/2011.07176} {arXiv:2011.07176 [astro-ph.HE]}
  \BibitemShut {NoStop}%
\bibitem [{\citenamefont {Fern\'andez}\ and\ \citenamefont
  {Metzger}(2013)}]{Fernandez:2013tya}%
  \BibitemOpen
  \bibfield  {author} {\bibinfo {author} {\bibfnamefont {R.}~\bibnamefont
  {Fern\'andez}}\ and\ \bibinfo {author} {\bibfnamefont {B.~D.}\ \bibnamefont
  {Metzger}},\ }\bibfield  {title} {\bibinfo {title} {{Delayed outflows from
  black hole accretion tori following neutron star binary coalescence}},\
  }\href {https://doi.org/10.1093/mnras/stt1312} {\bibfield  {journal}
  {\bibinfo  {journal} {Mon. Not. Roy. Astron. Soc.}\ }\textbf {\bibinfo
  {volume} {435}},\ \bibinfo {pages} {502} (\bibinfo {year} {2013})},\ \Eprint
  {https://arxiv.org/abs/1304.6720} {arXiv:1304.6720 [astro-ph.HE]}
  \BibitemShut {NoStop}%
\bibitem [{\citenamefont {Metzger}\ \emph {et~al.}(2008)\citenamefont
  {Metzger}, \citenamefont {Thompson},\ and\ \citenamefont
  {Quataert}}]{Metzger:2007kj}%
  \BibitemOpen
  \bibfield  {author} {\bibinfo {author} {\bibfnamefont {B.~D.}\ \bibnamefont
  {Metzger}}, \bibinfo {author} {\bibfnamefont {T.~A.}\ \bibnamefont
  {Thompson}},\ and\ \bibinfo {author} {\bibfnamefont {E.}~\bibnamefont
  {Quataert}},\ }\bibfield  {title} {\bibinfo {title} {{On the Conditions for
  Neutron-Rich Gamma-Ray Burst Outflows}},\ }\href
  {https://doi.org/10.1086/526418} {\bibfield  {journal} {\bibinfo  {journal}
  {Astrophys. J.}\ }\textbf {\bibinfo {volume} {676}},\ \bibinfo {pages} {1130}
  (\bibinfo {year} {2008})},\ \Eprint {https://arxiv.org/abs/0708.3395}
  {arXiv:0708.3395 [astro-ph]} \BibitemShut {NoStop}%
\bibitem [{\citenamefont {{Wanajo}}\ \emph {et~al.}(2014)\citenamefont
  {{Wanajo}}, \citenamefont {{Sekiguchi}}, \citenamefont {{Nishimura}},
  \citenamefont {{Kiuchi}}, \citenamefont {{Kyutoku}},\ and\ \citenamefont
  {{Shibata}}}]{Wanajo:2014}%
  \BibitemOpen
  \bibfield  {author} {\bibinfo {author} {\bibfnamefont {S.}~\bibnamefont
  {{Wanajo}}}, \bibinfo {author} {\bibfnamefont {Y.}~\bibnamefont
  {{Sekiguchi}}}, \bibinfo {author} {\bibfnamefont {N.}~\bibnamefont
  {{Nishimura}}}, \bibinfo {author} {\bibfnamefont {K.}~\bibnamefont
  {{Kiuchi}}}, \bibinfo {author} {\bibfnamefont {K.}~\bibnamefont
  {{Kyutoku}}},\ and\ \bibinfo {author} {\bibfnamefont {M.}~\bibnamefont
  {{Shibata}}},\ }\bibfield  {title} {\bibinfo {title} {{Production of All the
  r-process Nuclides in the Dynamical Ejecta of Neutron Star Mergers}},\ }\href
  {https://doi.org/10.1088/2041-8205/789/2/L39} {\bibfield  {journal} {\bibinfo
   {journal} {Astroph.J.Lett.}\ }\textbf {\bibinfo {volume} {789}},\ \bibinfo
  {eid} {L39} (\bibinfo {year} {2014})},\ \Eprint
  {https://arxiv.org/abs/1402.7317} {arXiv:1402.7317 [astro-ph.SR]}
  \BibitemShut {NoStop}%
\bibitem [{\citenamefont {{Qian}}\ \emph {et~al.}(1993)\citenamefont {{Qian}},
  \citenamefont {{Fuller}}, \citenamefont {{Mathews}}, \citenamefont {{Mayle}},
  \citenamefont {{Wilson}},\ and\ \citenamefont
  {{Woosley}}}]{1993PhRvL..71.1965Q}%
  \BibitemOpen
  \bibfield  {author} {\bibinfo {author} {\bibfnamefont {Y.-Z.}\ \bibnamefont
  {{Qian}}}, \bibinfo {author} {\bibfnamefont {G.~M.}\ \bibnamefont
  {{Fuller}}}, \bibinfo {author} {\bibfnamefont {G.~J.}\ \bibnamefont
  {{Mathews}}}, \bibinfo {author} {\bibfnamefont {R.~W.}\ \bibnamefont
  {{Mayle}}}, \bibinfo {author} {\bibfnamefont {J.~R.}\ \bibnamefont
  {{Wilson}}},\ and\ \bibinfo {author} {\bibfnamefont {S.~E.}\ \bibnamefont
  {{Woosley}}},\ }\bibfield  {title} {\bibinfo {title} {{Connection between
  flavor-mixing of cosmologically significant neutrinos and heavy element
  nucleosynthesis in supernovae}},\ }\href
  {https://doi.org/10.1103/PhysRevLett.71.1965} {\bibfield  {journal} {\bibinfo
   {journal} {Phys. Rev. Letters}\ }\textbf {\bibinfo {volume} {71}},\ \bibinfo
  {pages} {1965} (\bibinfo {year} {1993})}\BibitemShut {NoStop}%
\bibitem [{\citenamefont {Foucart}\ \emph {et~al.}(2020)\citenamefont
  {Foucart}, \citenamefont {Duez}, \citenamefont {Hebert}, \citenamefont
  {Kidder}, \citenamefont {Pfeiffer},\ and\ \citenamefont
  {Scheel}}]{Foucart:2020qjb}%
  \BibitemOpen
  \bibfield  {author} {\bibinfo {author} {\bibfnamefont {F.}~\bibnamefont
  {Foucart}}, \bibinfo {author} {\bibfnamefont {M.~D.}\ \bibnamefont {Duez}},
  \bibinfo {author} {\bibfnamefont {F.}~\bibnamefont {Hebert}}, \bibinfo
  {author} {\bibfnamefont {L.~E.}\ \bibnamefont {Kidder}}, \bibinfo {author}
  {\bibfnamefont {H.~P.}\ \bibnamefont {Pfeiffer}},\ and\ \bibinfo {author}
  {\bibfnamefont {M.~A.}\ \bibnamefont {Scheel}},\ }\bibfield  {title}
  {\bibinfo {title} {{Monte-Carlo neutrino transport in neutron star merger
  simulations}},\ }\href {https://doi.org/10.3847/2041-8213/abbb87} {\bibfield
  {journal} {\bibinfo  {journal} {Astrophys. J. Lett.}\ }\textbf {\bibinfo
  {volume} {902}},\ \bibinfo {pages} {L27} (\bibinfo {year} {2020})},\ \Eprint
  {https://arxiv.org/abs/2008.08089} {arXiv:2008.08089 [astro-ph.HE]}
  \BibitemShut {NoStop}%
\bibitem [{\citenamefont {Miller}\ \emph {et~al.}(2019)\citenamefont {Miller},
  \citenamefont {Ryan}, \citenamefont {Dolence}, \citenamefont {Burrows},
  \citenamefont {Fontes}, \citenamefont {Fryer}, \citenamefont {Korobkin},
  \citenamefont {Lippuner}, \citenamefont {Mumpower},\ and\ \citenamefont
  {Wollaeger}}]{Miller:2019dpt}%
  \BibitemOpen
  \bibfield  {author} {\bibinfo {author} {\bibfnamefont {J.~M.}\ \bibnamefont
  {Miller}}, \bibinfo {author} {\bibfnamefont {B.~R.}\ \bibnamefont {Ryan}},
  \bibinfo {author} {\bibfnamefont {J.~C.}\ \bibnamefont {Dolence}}, \bibinfo
  {author} {\bibfnamefont {A.}~\bibnamefont {Burrows}}, \bibinfo {author}
  {\bibfnamefont {C.~J.}\ \bibnamefont {Fontes}}, \bibinfo {author}
  {\bibfnamefont {C.~L.}\ \bibnamefont {Fryer}}, \bibinfo {author}
  {\bibfnamefont {O.}~\bibnamefont {Korobkin}}, \bibinfo {author}
  {\bibfnamefont {J.}~\bibnamefont {Lippuner}}, \bibinfo {author}
  {\bibfnamefont {M.~R.~.}\ \bibnamefont {Mumpower}},\ and\ \bibinfo {author}
  {\bibfnamefont {R.~T.}\ \bibnamefont {Wollaeger}},\ }\bibfield  {title}
  {\bibinfo {title} {{Full Transport Model of GW170817-Like Disk Produces a
  Blue Kilonova}},\ }\href {https://doi.org/10.1103/PhysRevD.100.023008}
  {\bibfield  {journal} {\bibinfo  {journal} {Phys. Rev. D}\ }\textbf {\bibinfo
  {volume} {100}},\ \bibinfo {pages} {023008} (\bibinfo {year} {2019})},\
  \Eprint {https://arxiv.org/abs/1905.07477} {arXiv:1905.07477 [astro-ph.HE]}
  \BibitemShut {NoStop}%
\bibitem [{\citenamefont {Hayashi}\ \emph {et~al.}(2022)\citenamefont
  {Hayashi}, \citenamefont {Fujibayashi}, \citenamefont {Kiuchi}, \citenamefont
  {Kyutoku}, \citenamefont {Sekiguchi},\ and\ \citenamefont
  {Shibata}}]{Hayashi:2021oxy}%
  \BibitemOpen
  \bibfield  {author} {\bibinfo {author} {\bibfnamefont {K.}~\bibnamefont
  {Hayashi}}, \bibinfo {author} {\bibfnamefont {S.}~\bibnamefont
  {Fujibayashi}}, \bibinfo {author} {\bibfnamefont {K.}~\bibnamefont {Kiuchi}},
  \bibinfo {author} {\bibfnamefont {K.}~\bibnamefont {Kyutoku}}, \bibinfo
  {author} {\bibfnamefont {Y.}~\bibnamefont {Sekiguchi}},\ and\ \bibinfo
  {author} {\bibfnamefont {M.}~\bibnamefont {Shibata}},\ }\bibfield  {title}
  {\bibinfo {title} {{General-relativistic neutrino-radiation
  magnetohydrodynamic simulation of seconds-long black hole-neutron star
  mergers}},\ }\href {https://doi.org/10.1103/PhysRevD.106.023008} {\bibfield
  {journal} {\bibinfo  {journal} {Phys. Rev. D}\ }\textbf {\bibinfo {volume}
  {106}},\ \bibinfo {pages} {023008} (\bibinfo {year} {2022})},\ \Eprint
  {https://arxiv.org/abs/2111.04621} {arXiv:2111.04621 [astro-ph.HE]}
  \BibitemShut {NoStop}%
\bibitem [{\citenamefont {Just}\ \emph {et~al.}(2016)\citenamefont {Just},
  \citenamefont {Obergaulinger}, \citenamefont {Janka}, \citenamefont
  {Bauswein},\ and\ \citenamefont {Schwarz}}]{Just:2015dba}%
  \BibitemOpen
  \bibfield  {author} {\bibinfo {author} {\bibfnamefont {O.}~\bibnamefont
  {Just}}, \bibinfo {author} {\bibfnamefont {M.}~\bibnamefont {Obergaulinger}},
  \bibinfo {author} {\bibfnamefont {H.~T.}\ \bibnamefont {Janka}}, \bibinfo
  {author} {\bibfnamefont {A.}~\bibnamefont {Bauswein}},\ and\ \bibinfo
  {author} {\bibfnamefont {N.}~\bibnamefont {Schwarz}},\ }\bibfield  {title}
  {\bibinfo {title} {{Neutron-star merger ejecta as obstacles to
  neutrino-powered jets of gamma-ray bursts}},\ }\href
  {https://doi.org/10.3847/2041-8205/816/2/L30} {\bibfield  {journal} {\bibinfo
   {journal} {Astrophys. J. Lett.}\ }\textbf {\bibinfo {volume} {816}},\
  \bibinfo {pages} {L30} (\bibinfo {year} {2016})},\ \Eprint
  {https://arxiv.org/abs/1510.04288} {arXiv:1510.04288 [astro-ph.HE]}
  \BibitemShut {NoStop}%
\bibitem [{\citenamefont {Eichler}\ \emph {et~al.}(1989)\citenamefont
  {Eichler}, \citenamefont {Livio}, \citenamefont {Piran},\ and\ \citenamefont
  {Schramm}}]{Eichler:1989ve}%
  \BibitemOpen
  \bibfield  {author} {\bibinfo {author} {\bibfnamefont {D.}~\bibnamefont
  {Eichler}}, \bibinfo {author} {\bibfnamefont {M.}~\bibnamefont {Livio}},
  \bibinfo {author} {\bibfnamefont {T.}~\bibnamefont {Piran}},\ and\ \bibinfo
  {author} {\bibfnamefont {D.~N.}\ \bibnamefont {Schramm}},\ }\bibfield
  {title} {\bibinfo {title} {{Nucleosynthesis, Neutrino Bursts and Gamma-Rays
  from Coalescing Neutron Stars}},\ }\href {https://doi.org/10.1038/340126a0}
  {\bibfield  {journal} {\bibinfo  {journal} {Nature}\ }\textbf {\bibinfo
  {volume} {340}},\ \bibinfo {pages} {126} (\bibinfo {year}
  {1989})}\BibitemShut {NoStop}%
\bibitem [{\citenamefont {Fujibayashi}\ \emph {et~al.}(2017)\citenamefont
  {Fujibayashi}, \citenamefont {Sekiguchi}, \citenamefont {Kiuchi},\ and\
  \citenamefont {Shibata}}]{Fujibayashi:2017xsz}%
  \BibitemOpen
  \bibfield  {author} {\bibinfo {author} {\bibfnamefont {S.}~\bibnamefont
  {Fujibayashi}}, \bibinfo {author} {\bibfnamefont {Y.}~\bibnamefont
  {Sekiguchi}}, \bibinfo {author} {\bibfnamefont {K.}~\bibnamefont {Kiuchi}},\
  and\ \bibinfo {author} {\bibfnamefont {M.}~\bibnamefont {Shibata}},\
  }\bibfield  {title} {\bibinfo {title} {{Properties of Neutrino-driven Ejecta
  from the Remnant of a Binary Neutron Star Merger: Pure Radiation
  Hydrodynamics Case}},\ }\href {https://doi.org/10.3847/1538-4357/aa8039}
  {\bibfield  {journal} {\bibinfo  {journal} {Astrophys. J.}\ }\textbf
  {\bibinfo {volume} {846}},\ \bibinfo {pages} {114} (\bibinfo {year}
  {2017})},\ \Eprint {https://arxiv.org/abs/1703.10191} {arXiv:1703.10191
  [astro-ph.HE]} \BibitemShut {NoStop}%
\bibitem [{\citenamefont {Kawaguchi}\ \emph {et~al.}(2023)\citenamefont
  {Kawaguchi}, \citenamefont {Fujibayashi},\ and\ \citenamefont
  {Shibata}}]{Kawaguchi:2022tae}%
  \BibitemOpen
  \bibfield  {author} {\bibinfo {author} {\bibfnamefont {K.}~\bibnamefont
  {Kawaguchi}}, \bibinfo {author} {\bibfnamefont {S.}~\bibnamefont
  {Fujibayashi}},\ and\ \bibinfo {author} {\bibfnamefont {M.}~\bibnamefont
  {Shibata}},\ }\bibfield  {title} {\bibinfo {title}
  {{Monte~Carlo\textendash{}based relativistic radiation hydrodynamics code
  with a higher-order scheme}},\ }\href
  {https://doi.org/10.1103/PhysRevD.107.023026} {\bibfield  {journal} {\bibinfo
   {journal} {Phys. Rev. D}\ }\textbf {\bibinfo {volume} {107}},\ \bibinfo
  {pages} {023026} (\bibinfo {year} {2023})},\ \Eprint
  {https://arxiv.org/abs/2209.12472} {arXiv:2209.12472 [astro-ph.HE]}
  \BibitemShut {NoStop}%
\bibitem [{\citenamefont {Izquierdo}\ \emph {et~al.}(2024)\citenamefont
  {Izquierdo}, \citenamefont {Abalos},\ and\ \citenamefont
  {Palenzuela}}]{Izquierdo:2023fub}%
  \BibitemOpen
  \bibfield  {author} {\bibinfo {author} {\bibfnamefont {M.~R.}\ \bibnamefont
  {Izquierdo}}, \bibinfo {author} {\bibfnamefont {J.~F.}\ \bibnamefont
  {Abalos}},\ and\ \bibinfo {author} {\bibfnamefont {C.}~\bibnamefont
  {Palenzuela}},\ }\bibfield  {title} {\bibinfo {title} {{Guided moments
  formalism: A new efficient full-neutrino treatment for astrophysical
  simulations}},\ }\href {https://doi.org/10.1103/PhysRevD.109.043044}
  {\bibfield  {journal} {\bibinfo  {journal} {Phys. Rev. D}\ }\textbf {\bibinfo
  {volume} {109}},\ \bibinfo {pages} {043044} (\bibinfo {year} {2024})},\
  \Eprint {https://arxiv.org/abs/2312.09275} {arXiv:2312.09275 [astro-ph.HE]}
  \BibitemShut {NoStop}%
\bibitem [{\citenamefont {Maggiore}\ \emph {et~al.}(2020)\citenamefont
  {Maggiore} \emph {et~al.}}]{Maggiore:2019uih}%
  \BibitemOpen
  \bibfield  {author} {\bibinfo {author} {\bibfnamefont {M.}~\bibnamefont
  {Maggiore}} \emph {et~al.},\ }\bibfield  {title} {\bibinfo {title} {{Science
  Case for the Einstein Telescope}},\ }\href
  {https://doi.org/10.1088/1475-7516/2020/03/050} {\bibfield  {journal}
  {\bibinfo  {journal} {JCAP}\ }\textbf {\bibinfo {volume} {03}},\ \bibinfo
  {pages} {050}},\ \Eprint {https://arxiv.org/abs/1912.02622} {arXiv:1912.02622
  [astro-ph.CO]} \BibitemShut {NoStop}%
\bibitem [{\citenamefont {Evans}\ \emph {et~al.}(2023)\citenamefont {Evans}
  \emph {et~al.}}]{Evans:2023euw}%
  \BibitemOpen
  \bibfield  {author} {\bibinfo {author} {\bibfnamefont {M.}~\bibnamefont
  {Evans}} \emph {et~al.},\ }\bibfield  {title} {\bibinfo {title} {{Cosmic
  Explorer: A Submission to the NSF MPSAC ngGW Subcommittee}},\ }\href@noop {}
  {\bibfield  {journal} {\bibinfo  {journal} {arXiv}\ } (\bibinfo {year}
  {2023})},\ \Eprint {https://arxiv.org/abs/2306.13745} {arXiv:2306.13745
  [astro-ph.IM]} \BibitemShut {NoStop}%
\end{thebibliography}%

\end{document}